\documentstyle[amssymb,graphicx,twoside,submitp]{pio}

\def\RB{{\Large $\mathbf{+}$}}
\def\DB{{\Large $\mathbf{\times}$}}
\def\ket#1{| #1 \rangle}

\def\bm#1{\mbox{\boldmath $#1$}}

\begin{document}

\title{QUANTUM CRYPTOGRAPHY}

\author{Miloslav Du\v{s}ek \\
{\it Department of Optics, Palack\'{y} University\\
17. listopadu 50, 772\,00 Olomouc, Czech Republic}\\ \and
Norbert L\"{u}tkenhaus \\
{\it Institut f\"{u}r Optik, Information und Photonik \\
Universit\"{a}t Erlangen-N\"{u}rnberg \\
Staudtstr. 7/B3, 91058 Erlangen, Germany}\\ \and
Martin Hendrych \\
{\it  ICFO - Institut de Ci\`{e}ncies Fot\`{o}niques \\
Parc Mediterrani de la Tecnologia, Avda. Canal Ol\'{\i}mpic, s/n \\
08860 Castelldefels (Barcelona), Spain}}

\maketitle

\tableofcontents

\section{Ciphering}

\subsection{Introduction, cryptographic tasks}

There is no doubt that electronic communications have become one
of the main pillars of the modern society and their ongoing boom
requires the development of new methods and techniques to secure
data transmission and data storage. This is the goal of
cryptography. Etymologically derived from Greek
$\kappa\rho\upsilon\pi\tau\acute{o}\varsigma$,
hidden or secret, and
$\gamma\rho\alpha\varphi\acute{\eta}$, writing, cryptography may
generally be defined as the art of writing (encryption) and
deciphering (decryption) messages in code in order to ensure their
confidentiality, authenticity, integrity and non-repudiation.
Cryptography and cryptanalysis, the art of codebreaking, together
constitute cryptology ($\lambda\acute{o}\gamma o\varsigma$, a word).

Nowadays many paper-based communications have already been
replaced by electronic means, raising the challenge to find
electronic counterparts to stamps, seals and hand-written
signatures. The growing variety of applications brings many tasks
that must be solved. Let us name a few. The fundamental task of
cryptography is to allow two users to render their communications
unintelligible to any third party, while for the two legitimate
users the messages remain intelligible. The goal of identification
is to verify the identities of the communicating parties. Another
cryptographic task is secret sharing: A secret, e.g., a password,
is split into several pieces in such a way that when a certain
minimal subset of the pieces is put together, the secret is
recovered. Other cryptographic applications are, for example,
digital signatures, authentication of messages, zero-knowledge
proofs, and so on.

At all times people have wished to have the possibility to
communicate in secrecy so as to allow nobody to overhear their
messages. Archeological excavations have revealed that various
types of cryptography had already been used by ancient
civilizations in Mesopotamia, India, or China
(\cite{Codebreakers}). Four thousand years ago, ancient Egyptians
used modified hieroglyphs to conceal their messages. In the Iliad,
Homer depicts how Proetus, the king of Argolis, sends Bellerophon
to Lycia with ``a lethal message, coded symbols inscribed on a
folded tablet" (\cite{Iliad1}).

In the $5^{\rm th}$ century BC, the Spartans in Greece designed
the \emph{Skytale} cryptodevice, based on transposition of letters
(\cite{Sparta}). A stripe of parchment or leather was wound around
a wooden baton, across which the message was written. When the end
of line was reached, the baton was rotated. After the parchment
was unwrapped, the letters looked scrambled and only the person
who possessed a baton of an identical shape could recover the
message.

Another favorite and easy cipher is the substitution cipher, which
substitutes each letter of a message with another letter, number
or a symbol. An example is the Caesar cipher (\cite{Stinson}). To
communicate between the Roman legions scattered over the Roman
republic, Gaius Julius Caesar used a cipher, where each letter of
a message was advanced by three letters in the alphabet; A was
replaced by D, B was replaced by E, C by F, and so on. Similar
substitution cipher is also described in Kama Sutra.

During the Middle Ages, most cryptosystems were based on
transposition or substitution or a combination of both
(\cite{Leary}). However, neither of these ciphers is secure,
because it is possible to break them exploiting various
characteristic properties of the language, such as the frequency
of individual letters and their clusters.

The invention of the telegraph in the 1830s enormously facilitated
communications between people. This ancestor of modern
communications, however, had a serious drawback from the
cryptographic point of view -- the content of the transmitted
message was known to the telegraph operator. As a consequence,
various codebooks were designed by people and companies that
wanted to keep their communications private. The codebooks
translated significant words and phrases into short, nonsensical
words. The codes served two purposes: first, they reduced the size
of the message and thus decreased the costs because telegrams were
charged per transmitted character; and second, if the codebook was
kept secret, the codes became a cipher.

The two world wars of the $20^{\rm th}$ century accelerated
the development of new cryptographic techniques.
Cryptographers tried to design a system where the
encryption and decryption algorithms could be publicly
known, but the secrecy of the message would be guaranteed
by some secret information, the cryptographic key, shared
between the users.  In 1917, Gilbert S. Vernam proposed an
unbreakable cryptosystem, hence called the Vernam cipher or
One-time Pad (\cite{Vernam}). Its unconditional security
has been proved by Claude E. Shannon (in terms of
information theory) in 1949 (\cite{Shannon}). The One-time
Pad is a special case of the substitution cipher, where
each letter is advanced by a random number of positions in
the alphabet. These random numbers then form the
cryptographic key that must be shared between the sender
and the recipient. Even though the Vernam cipher offers
unconditional security against adversaries possessing
unlimited computational power and technological abilities,
it faces the problem of how to securely distribute the key.
That is why it did not become widespread as Vernam had
hoped. On the other hand, there are many military and
diplomatic applications, where the security of
communications outweighs the severe key management
problems. The Vernam cipher was used by the infamous spies
Theodore A. Hall, Klaus Fuchs, the Rosenbergs and others,
who were passing atomic secrets to Moscow. Che Guevara also
encrypted his messages to Fidel Castro by means of the
One-time Pad. It was employed in securing the hot line
between Washington and Moscow and it is said to be used for
communications between nuclear submarines and for some
embassy communications. We will come back to the Vernam
cipher later on, as it is this cipher that is very
expedient for quantum key distribution.

In 1918, Arthur Scherbius invented an ingenious electric cipher
machine, called Enigma, which was patented a year later
(\cite{Deavours}). The Enigma consisted of a set of rotating wired
wheels, which performed a very sophisticated substitution cipher.
After various improvements, it was adopted by the German Navy in
1926, the German Army in 1928, and the Air Force in 1935, and it
was used by the Germans and Italians throughout World War II. The
military Enigma had incredible $159 \times 10^{18}$  possible
settings (cryptographic keys). The immense number of potential
keys led Alan Turing to construct the first electronic computer,
which helped break the Enigma ciphers in the course of the War.
Today a Pentium-based computer can unscramble an Enigma-encrypted
message within minutes.

\subsection[Asymmetrical ciphers]{Asymmetrical ciphers (Public-key cryptography)}

A new surge of interest in cryptography was triggered by the
upswing in electronic communications in the late 70s of the
$20^{\rm th}$ century. It was essential to enable secure
communication between users who have never met before and share no
secret cryptographic key. The question was how to distribute
the key in a secure way. The solution was found by Whitfield
Diffie and Martin E. Hellman, who invented public-key cryptography
in 1976 (\cite{Diffie}). The ease of use of public-key
cryptography, in turn, stimulated the boom of electronic commerce
during the 1990s. Notice, however, that asymmetric ciphers can
provide users who have never met with a \emph{secret} channel but
-- without the help of a Trusted Authority -- it cannot prove the
\emph{identity} of users.

Public-key cryptography requires two keys -- the public key and
the private key, which form a key pair. The recipient
generates two keys, makes the public key public
and keeps his private key in a secret place to ensure
its private possession. The algorithm is designed in such a way
that anyone can encrypt a message using the public key, however,
only the legitimate recipient can decrypt the message using
his/her private key.

Of course, there is a problem of authenticity of the public key.
Therefore public keys are distributed through Trusted Authorities
in practice.

The security of public-key cryptography rests on various
computational problems, which are believed to be intractable. The
encryption and decryption algorithms utilize the so-called one-way
functions. One-way functions are mathematical functions that are
easy to compute in one direction, but their inversion is very
difficult (by ``difficult" it is meant that the number of the
required elementary operations increases exponentially with the
length of the input number). It is, e.g., very easy to multiply
two prime numbers, but to factor the product of two large primes
is already a difficult task. Other public-key cryptosystems are
based, e.g., on the difficulty of the discrete logarithm problem
in Abelian groups on elliptic curves or other finite groups.
However, it is important to point out that no ``one-way function"
has been proved to be one-way; they are merely believed to be.
\footnote{This believe is based on the experience that even years
of effort of many experts do not proof the opposite.} Public-key
cryptography cannot provide unconditional security. We speak about
computational security.

Today the most widely used public-key system is the RSA
cryptosystem. RSA was invented in 1977 by Ronald Rivest, Adi
Shamir and Leonard Adleman (\cite{Rives}), whose names form the
acronym. RSA exploits the difficulty of factoring large numbers.
The receiver picks two large primes $p$ and $q$ and makes their
product public. Further, he chooses two large natural numbers $d$
and $e$ [such that $(de-1)$ is divisible by $(p-1)(q-1)$]. The
product $pq$ together with the number $e$ constitutes the public
key. Using this key, anyone can encrypt a message $P$ ($P<pq$)
employing a simple algorithm: $C = P^e \bmod pq$, where $C$ is the
resulting cipher text. The cipher text can easily be decrypt if
the private key $d$ is known: $P = C^d \bmod pq$.  However, in
order to invert the algorithm without knowing the private key $d$
it is necessary to find the prime factors of the modulus. Although
there are several other ways to attack the RSA system, the most
promising one still seems to be to attempt to factor the modulus.

In 1976 Richard Guy wrote (\cite{Guy}): ``I shall be surprised if
anyone regularly factors numbers of size $10^{80}$ without special
form during the present century". The first challenge to break a
425-bit RSA key (equivalent to 129 decimal digits) was published
in Scientific American in 1977 (\cite{Gardner}). Ronald Rivest
calculated that to factor a 125-digit number, the product of two
63-digit primes, would take at least $40 \times 10^{15}$ years
(about one million times the age of the universe) with the best
factoring algorithms then known. However, 17 years later, in 1994,
new factoring algorithms had been discovered and computer power
had advanced to such a level that it took 1600 computers (and two
fax machines!) interconnected over the Internet only 8 months.
Today a single Pentium-based PC could do the same job.

While breaking 425-bit RSA required a large number of computers,
in February 1999 it was only 185 machines that managed to factor a
465-bit RSA modulus in 9 weeks. At that time, 95\% of e-commerce
on the Internet was protected by 512-bit keys (155-digit number).
A 512-bit number was factored in August 1999 by 292 machines. That
means that neither 512-bit keys provide sufficient security for
anything more than very short-term security needs. All these
challenges have served to estimate the amount of work and the cost
of breaking a key of a certain size by public efforts. It is
obviously much more difficult to estimate what can be achieved by
private and governmental efforts with much larger budgets.

A network of computers is not the only way to factor large
integers. In 1999 Adi Shamir proposed the TWINKLE device
(\cite{Shamir}) -- a massively parallel optoelectronic factoring
device, which is about three orders of magnitude faster than a
conventional fast PC and can facilitate the factoring of 512- and
768-bit keys. Today it is already recommended to move to longer
key lengths and to use key sizes of 2048 bits for corporate use
and 4096 bits for valuable keys.

Another menace to the security of public-key cryptography could
originate from the construction of a quantum computer. The
decryption using a quantum computer would take about the same time
as the encryption, thereby making public-key cryptography
worthless. Algorithms capable of doing so have already been
developed (\cite{Shor}) and first experiments with small-scale
quantum computers successfully pave the way to more sophisticated
devices (\cite{Vandersypen}).

\subsection{Symmetrical ciphers (Secret-Key Cryptography)}

In secret-key cryptography users must share a secret key
beforehand. The common key is then used for both encryption and
decryption.\footnote{Secret-key cryptography can provide its users
even with unconditional security if they share a sufficiently long
key (using Vernam cipher). But symmetric algorithms with the key
shorter than the message are not unconditionally secure.} Secure
key distribution is the main drawback of secret-key cryptosystems.
The security of communications is reduced to the security of
secret-key distribution. In order to avoid the necessity of
personal meetings or courier services to exchange the secret key,
some users use public-key cryptography to distribute the key,
which is then used in a secret-key cryptosystem. In such a case,
even if the symmetric cipher was unconditionally secure the
security of the whole system will be degraded to computational
security. These so-called hybrid systems have gained a widespread
use, because they combine the speed of secret-key systems with the
efficiency of key management of public-key systems. They have been
used for electronic purchases, financial transactions, ATM
transactions and PIN encryptions, identification and
authentication of cellular phone conversations, electronic
signatures, and many other applications, whose number is swelling.

The most spread secret-key cryptosystem is the Data Encryption
Standard (DES) and its variations. Due to its frequent use in the
hybrid systems, it is the most often used cryptosystem ever. DES
was developed by IBM and the U.S. government in 1975 and it was
adopted as a standard two years later. DES is an example of a
block cipher -- an algorithm that takes a fixed-length string of
plaintext and transforms it through a series of operations into
another ciphertext of the same length. In the case of DES, the
block size is 64 bits. The transformation depends on the key. The
algorithm consists of the cascade of 16 iterations of
substitutions and transpositions and can easily be implemented in
hardware, where it can reach very high speeds of encryption.

DES has experienced a similar wave of attacks as public-key
cryptosystems. The algorithm uses a 56-bit key, which is reused to
encrypt the entire message. As a consequence, it is only
computationally secure. In 1997, RSA Data Security, Inc. published
their first challenge to decrypt a plaintext message scrambled by
DES. It took 96 days to break it. The researchers applied ``brute
force" by searching the entire keyspace of $2^{56}$ possible keys
on a large number of computers (\cite{Wiener}). In January 1998, a
new prize was offered. The winner of the contest used the idle
time of computers connected to the Internet. More than 50,000 CPUs
were linked together. The key was found after 41 days
(\cite{DESCrack}). Another group of codebreakers chose a different
approach. They built a single machine, which revealed the
encrypted message ``It's time for those 128-, 192-, and 256-bit
keys" after only 56 hours, searching at a rate of 88 billion keys
per second (\cite{DESCracker}).

In the challenge in January 1999, the two previous winners
combined their efforts to find the key in only 22 hours and 15
minutes, testing 245 billion keys per second. In 1993, Michael
Wiener designed a DES key search machine which, based on 1997's
technology, would break DES in 3.5 hours (\cite{Wiener}). The same
machine based on 2000's technology would take only 100 seconds
(\cite{Silverman}). The exhaustive search is not the only possible
attack on DES. During the 1990s, other successful attacks were
proposed that exploit the internal structure of the cipher
(\cite{Biham}).

Cryptographers attempted to improve the security of DES. Triple
DES, DESX and other modifications were developed. In October 2000,
a four-year effort to replace the aging DES culminated in the
announcement of a new standard, the Advanced Encryption Standard
(\cite{AES}). It uses blocks of 128 bits and key sizes of 128,
192, and 256 bits. This standard was approved in December 2001 and
went into effect in May 2002. How long will it last?

In summary, the security of conventional techniques relies on the
assumption of limited advancement of mathematical algorithms and
computational power in the foreseeable future, and also on limited
financial resources available to a potential adversary.
Computationally secure cryptosystems, no matter whether public- or
secret-key, will always be threatened by breakthroughs, which are
difficult to predict, and even steady progress of code-breaking
allows the adversary to ``reach back in time" and break older,
earlier captured, communications encrypted with weaker keys.

Another common problem of conventional cryptographic methods is
the so-called side-channel cryptanalysis (\cite{Rosa}). Side
channels are undesirable ways through which information related to
the activity of the cryptographic device can leak out. The attacks
based on side-channel information do not assault the mathematical
structure of cryptosystems, but their particular implementations.
It is possible to gain information by measuring the amount of time
needed to perform some operation, by measuring power consumption,
heat radiation or electromagnetic emanation. The problem of side
channels will be further discussed in Section~\ref{sec_side_ch}.

\subsection{Vernam Cipher, key distribution problem}

Classical cryptography can provide an unbreakable cipher, which
resists adversaries with unlimited computational and technological
power -- the Vernam cipher. The Vernam cipher was invented in 1917
by the AT\&T engineer Gilbert S.~Vernam (\cite{Vernam}), who
thought it would become widely used for automatic encryption and
decryption of telegraph messages.

\begin{sloppypar}
The Vernam cipher belongs to the symmetric secret-key
ciphers, i.e., the same key is used for both, encryption
and decryption. The principle of the cipher is that if a
random key is added to a message, the bits of the resulting
string are also random and carry no information about the
message. If we use the binary logic, unlike Vernam who
worked with a 26-letter alphabet, the encryption algorithm
$E$ can be written as
\begin{equation}
E_K(M) = (M_1 + K_1,M_2+K_2, \dots, M_n+K_n) \bmod 2,
\end{equation}
where $M=(M1,M2, \dots, M_n) $ is the message to be
encrypted and $K=(K_1,K_2, \dots, K_n)$ is the key
consisting of random bits. The message and the key are
added bitwise modulo 2, or exclusive OR without carries.
The decryption $D$ of ciphertext $C=E_K(M)$ is identical to
encryption, because double modulo-2 addition is the identity,
therefore
\begin{equation}
M=D_K(C)=(C_1+K_1, C_2+K_2, \dots, C_n+K_n) \bmod 2.
\end{equation}
For this system to be unconditionally secure, three requirements
are imposed on the key: (1) The key must be as long as the
message; (2) it must be purely random; (3) it may be used only
once.\footnote{If a key $K$ is used twice to encode two different
messages $M$ and $M'$ into ciphertexts $C$ and $C'$ then one can
see that $(C_1+C'_1, C_2+C'_2, \dots, C_n+C'_n) \bmod 2 =
(M_1+M'_1, M_2+M'_2, \dots, M_n+M'_n) \bmod 2$.}
This was shown by Claude E. Shannon (\cite{Shannon}), who laid the
foundations of communication theory from the cryptographic point
of view and compared various cryptosystems with respect to their
secrecy. Until 1949 when his paper was published, the Vernam
cipher was considered unbreakable, but it was not mathematically
proved. If any of these requirements is not fulfilled, the
security of the system is jeopardized. A good example is the
revelation of the WWII atomic spies because of repetitive use of
the key incorrectly prepared by the KGB (\cite{KGB}).
\end{sloppypar}

The main drawback of the Vernam cipher is the necessity to
distribute a secret key as long as the message, which prevented it
from wider use. The cipher has so far found applications mostly in
the military and diplomatic services. It is here that quantum
mechanics comes in handy and readily offers a solution. Quantum mechanics
gives us the power to detect eavesdropping. Taking into account the problem
of authentication, that requires the communication parties to share a certain
amount of secret information, quantum cryptography provides a tool for an unlimited
secret-key growing.

\section{Quantum key distribution\label{QKD}}

\subsection{The principle, eavesdropping can be detected}

As mentioned above, the main problem of secret-key
cryptosystems is the secure distribution of keys. While the
security of classical cryptographic methods can be
undermined by advances in technology and mathematical
algorithms, the quantum approach can provide unconditional
security. The principle of quantum cryptography consists in
the use of non-orthogonal quantum states. Its security is
guaranteed by the Heisenberg uncertainty principle, which
does not allow us to discriminate non-orthogonal states
with certainty and without disturbing the measured system.

Within the framework of classical physics, it is impossible
to reveal potential eavesdropping, because information
encoded into any property of a classical object can be
acquired without affecting the state of the object. All
classical signals can be monitored passively. In classical
communications, one bit of information is encoded into two
distinguishable states of billions of photons, electrons,
atoms or other carriers. It is always possible to passively
listen in by splitting off part of the signal and performing a
measurement on it.

In quantum cryptosystems the inviolateness of the channel is
constantly tested by the use of non-orthogonal quantum states as
information carriers. Because information is encoded into states
with non-zero overlap, it cannot be read, copied or split without
introducing detectable disturbances.

It should be noted that quantum mechanics does not avert
eavesdropping; it only enables us to detect the presence of an
eavesdropper. Since only the cryptographic key is transmitted, no
information leak can take place when someone attempts to listen
in. When discrepancies are found, the key is simply discarded and
the users repeat the procedure to generate a new key.

\subsection{Quantum measurement\label{SectionQMeasurement}}

Measurement in quantum physics differs substantially from
the measurement in classical physics. According to quantum
theory any measurement can distinguish with certainty (i.e.
without errors or inconclusive results) only among specific
orthogonal state vectors (that form the so called measurement
basis). Non-orthogonal states cannot be distinguished
perfectly. Furthermore, quantum measurement disturbs the
system in general. If the system is in a state that
cannot be expressed as a multiple of one of the
measurement-basis vectors but only as their linear
superposition then this state is changed after the
measurement. The original state is ``forgotten'' during the
measurement process and randomly changed to the state
corresponding to one of the basis vectors. Right this is
the key feature of the quantum world that enables to detect
the eavesdropping. Eavesdropping is nothing else than a
kind of measurement on the information carrier. If
non-orthogonal states are used in transmission,
eavesdropping must disturb some of them, i.e. induce
errors.  With a suitably designed protocol, these errors
can later be discovered by the legitimate users of the
channel, as will be seen in Section \ref{BB84}.

\subsection{Quantum states cannot be cloned}
      \label{SectionCloning}

The linearity of quantum mechanics prohibits from cloning
arbitrary unknown quantum states (\cite{Wooters}). A device
intended to make a copy of, say, a photon with horizontal
polarization $|H\rangle$, needs to perform the following operation
\begin{equation}
|{\mathrm{copier}}_0\rangle |{\mathrm{blank}}\rangle
|H\rangle \rightarrow |{\mathrm{copier}}_1 \rangle
|H\rangle |H\rangle,
\end{equation}
and similarly for orthogonal vertical polarization
$|V\rangle$
\begin{equation}
|{\mathrm{copier}}_0\rangle |{\mathrm{blank}}\rangle
|V\rangle \rightarrow |{\mathrm{copier}}_2 \rangle
|V\rangle |V\rangle,
\end{equation}
where $|{\mathrm{copier}}_0\rangle$ is the initial state of
the copier, $|{\mathrm{copier}}_1\rangle$ and
$|{\mathrm{copier}}_2\rangle$ are its final states, and
$|{\mathrm{blank}}\rangle$ denotes the initial ``empty''
state of the ancillary system (photon) to which the
information (polarization state) should be copied. However,
if we want to copy a linear superposition of states
$|H\rangle$ and $|V\rangle$, we obtain
\begin{eqnarray}
|{\mathrm{copier}}_0\rangle |{\mathrm{blank}}\rangle
(\alpha\, |H\rangle + \beta\, |V\rangle) &\!\!=\!\!&
\alpha\, |{\mathrm{copier}}_0\rangle
|{\mathrm{blank}}\rangle |H\rangle + \beta\,
|{\mathrm{copier}}_0\rangle  |{\mathrm{blank}}\rangle
|V\rangle
\nonumber \\
 &\!\!\rightarrow\!\!& \alpha\, |{\mathrm{copier}}_1\rangle |H\rangle |H\rangle
 + \beta\, |{\mathrm{copier}}_2\rangle |V\rangle |V\rangle,
\end{eqnarray}
which is different from the required state
\begin{eqnarray}
&&|{\mathrm{copier}}_3\rangle \,
(\alpha\, |H\rangle + \beta\, |V\rangle)\,(\alpha\, |H\rangle +
\beta\, |V\rangle) \nonumber \\
&&= |{\mathrm{copier}}_3\rangle \left(
\alpha^2\, |H\rangle |H\rangle +
\alpha\beta\, |H\rangle |V\rangle + \beta\alpha\, |V\rangle|H\rangle +
\beta^2\, |V\rangle |V\rangle \right),
\end{eqnarray}
regardless of whether states $|{\mathrm{copier}}_1\rangle$
and $|{\mathrm{copier}}_2\rangle$ are identical
(and equal to $|{\mathrm{copier}}_3\rangle$) or not. The
unitarity of quantum evolution requires that
\begin{equation}
\langle H|V\rangle\, \langle
{\mathrm{blank}}|{\mathrm{blank}}\rangle\,
\langle{\mathrm{copier}}_0|{\mathrm{copier}}_0\rangle =
\langle H|V\rangle \, \langle H|V\rangle \, \langle
{\mathrm{copier}}_1|{\mathrm{copier}}_2\rangle,
\end{equation}
what can be satisfied only when the states to be copied are
orthogonal.

Thus, the general state of a quantum object cannot be copied
precisely. Duplicating can be done only approximately so that any
of the resulting states is not exactly equal to the original. An
optimal universal machine for approximate cloning of qubits was
first designed by \cite{Buzek96}.

\subsection{Protocol BB84\label{BB84}}

Quantum key distribution (QKD) was born in 1984 when
Charles H. Bennett and Gilles Brassard came up with an idea
of how to securely distribute a random cryptographic key
with the help of quantum mechanics (\cite{Bennett}). Hence,
the protocol is called BB84. Drawing upon Stephen Wiesner's
ideas about unforgeable quantum money (\cite{Wiesner},
original manuscript written circa 1969), Bennett and
Brassard presented a protocol that allows users to
establish an identical and purely random sequence of bits
at two different locations, while allowing to reveal any
eavesdropping with a very high probability.

The crucial point of the BB84 protocol is the use of two
conjugated bases. The sender of the message encodes logical zeros
and ones into two orthogonal states of a quantum system. But for
each bit she randomly changes this pair of states -- i.e., she
chooses one of two bases. Each state vector of one basis has
equal-length projections onto all vectors of the other basis. That
is, if a measurement on a system prepared in one basis is
performed in the other basis, its outcome is entirely random and
the system ``loses all the memory" of its previous state. In fact,
the non-orthogonal signal states are used for testing the
transmission channel -- checking it for eavesdropping.

We need not consider any particular quantum system.
However, in order to provide an example let us suppose that
information is encoded into polarization states of
individual quanta of light -- photons. One basis can
consist, e.g., of horizontal and vertical polarization
states of photons, $|H\rangle$ and $|V\rangle$, resp.; let
us call this basis \emph{rectilinear}. The other basis,
\emph{diagonal}, would consist of states of linear
polarizations at $45^{\circ}$ (anti-diagonal), $|A\rangle$,
and $135^{\circ}$ (diagonal), $|D\rangle$, whereas
\begin{eqnarray}
\label{diag} |A\rangle &=& \frac{1}{\sqrt{2}}(|H\rangle +
|V\rangle), \nonumber \\
|D\rangle &=& \frac{1}{\sqrt{2}}(|H\rangle - |V\rangle).
\end{eqnarray}
These four states satisfy the following relations
\begin{eqnarray}
\label{nonorthogonality}
&\langle H|V \rangle = \langle A|D \rangle = 0 \nonumber& \\
&\langle H|H \rangle = \langle V|V \rangle = \langle A|A \rangle
= \langle D|D \rangle = 1,& \\
&|\langle H|A \rangle|^2 = |\langle H|D \rangle|^2 = |\langle V|A
\rangle|^2 = |\langle V|D \rangle|^2 =1/2.&
\nonumber
\end{eqnarray}
Any measurement in the rectilinear (diagonal) basis on photons
prepared in the diagonal (rectilinear)  basis will yield random
outcomes with equal probabilities. On the other hand, measurements
performed in the basis identical to the basis of preparation of
states will produce deterministic results.\footnote{We could also
consider a third basis consisting of right and left circular
polarizations whose vectors satisfy relations analogous to
Eqs.~(\ref{nonorthogonality}). Any two of these three mentioned
bases suffice for secure quantum key distribution.}

At the beginning, the two parties that wish to communicate,
traditionally called Alice and Bob, agree that, e.g.,
$|H\rangle$ and $|A\rangle$ stand for the bit value ``0",
and $|V\rangle$ and $|D\rangle$ stand for a bit value ``1".
Now Alice, the sender, generates a sequence of random bits
that she wants to transmit, and randomly and independently
for each bit she chooses her encoding basis, rectilinear or
diagonal. Physically it means that she transmits photons in
the four polarization states $|H\rangle, |V\rangle,
|A\rangle,$ and $|D\rangle$ with equally distributed
frequencies. Bob, the receiver, randomly and independently
of Alice, chooses his measurement bases, either rectilinear
or diagonal. Statistically, their bases coincide in 50\% of
cases, when Bob's measurements provide deterministic
outcomes and perfectly agree with Alice's bits. In order to
know when the outcomes were deterministic, Alice and Bob
need an auxiliary public channel to tell each other what
basis they had used for each transmitted and detected
photon. This classical channel may be tapped, because it
transmits only information about the used bases, not about
the particular outcomes of the measurements. Whenever their
bases coincide, Alice and Bob keep the bit. On the other
hand, the bit is discarded when they chose different bases,
or Bob's detector failed to register a photon due to
imperfect efficiency of detectors or the photon was lost
somewhere on the way. Any potential eavesdropper,
traditionally called Eve, who listens into this
conversation can only learn whether they both set the
rectilinear or diagonal basis, but not whether Alice had
sent a ``0" or ``1". The protocol is represented in Table
\ref{tableBB84}.

\begin{table}
\begin{center}
\begin{tabular}{|c|c|c|c|c|c|c|c|}
\hline
 0 & 1 & 1 & 0 & 0 & 1 & 0 & 1 \\ \hline
\DB & \DB & \RB & \RB & \RB & \DB & \RB & \DB \\ \hline $\ket{A}$
& $\ket{D}$ & $\ket{V}$ & $\ket{H}$ &
$\ket{H}$ & $\ket{D}$ & $\ket{H}$ & $\ket{D}$ \\
\hline \hline
\DB & \RB & \RB & \DB & \DB & \RB & \RB & \DB \\
\hline
$\ket{A}$ & rand & $\ket{V}$ & rand & rand & rand & lost & $\ket{D}$ \\
\hline \hline \DB & \RB & \RB & \DB & \DB & \RB & -- & \DB \\
\hline OK & -- & OK & -- & -- & -- & -- & OK \\ \hline
0 & -- & 1 & -- & -- & -- & -- & 1 \\
\hline
\end{tabular}
\end{center}
\label{tableBB84} \caption{BB84 Protocol. $1^{\rm st}$ line --
Alice's random bits. $2^{\rm nd}$ line -- Alice's random
polarization bases; ``$+$" and ``$\times$" stand for the
rectilinear and diagonal bases, resp. $3^{\rm rd}$ line -- actual
polarization of transmitted photons. $4^{\rm th}$ line -- Bob's
random detection bases. $5^{\rm th}$ line -- polarization of
detected photons; `rand' stands for a random outcome. $6^{\rm th}$
line -- Bob publicly announces his measurement bases. $7^{\rm th}$
line -- Alice publicly replies when Bob set the correct
measurement basis. $8^{\rm th}$ line -- the cryptographic key.}
\end{table}

\subsection{Eavesdropping, intercept-resend attack}\label{Int_res_signal_basis}

If Eve is present and wants to eavesdrop on the channel, she
cannot passively monitor the transmissions (single quantum cannot
be split and its state cannot be copied without introducing
detectable disturbances, as discussed above). What Eve can do is
either to intercept the photons sent by Alice, perform
measurements on them and resend them to Bob or to attach some
probe to the signal photon, i.e., to let interact some  system in
her hands  with the quantum system carrying information, keep it
and measure it later. To understand the effect of eavesdropping we
will consider first only the intercept-resend attack. As Alice
alternates her encoding bases at random, Eve does not know the
basis to make a measurement in. She must choose her measurement
bases at random as well. Half the time she guesses right and she
resends correctly polarized photons. In 50\% of cases, though, she
measures in the wrong basis, which produces errors. For example,
let us suppose that Alice sends a ``1" in the rectilinear basis,
i.e., state $|V\rangle$, Eve measures in the diagonal basis, and
Bob measures in the rectilinear basis (otherwise the bit would be
discarded). Now, no matter whether Eve detects and resends
$|A\rangle$ or $|D\rangle$, Bob has a 50\% chance to get
$|H\rangle$, i.e., a binary ``0", instead of $|V\rangle$. Thus, if
we consider a continuous intercept-resend eavesdropping, Bob finds
on average errors in 25\% of those bits that he successfully
detects. If Alice and Bob agree to disclose part of their strings
in order to compare them, they can discover these errors. When
they set identical bases, their bit strings should be in perfect
agreement. When discrepancies are found, Eve is suspected of
tampering with the photons, and the cryptographic key is thrown
away. Thus, no information leakage occurs even in the case of
eavesdropping. If their strings are identical, the key is deemed
secure and secret,\footnote{The probability that eavesdropping
will not be detected decreases exponentially with the increasing
number of compared bits.} and can be used for the above-mentioned
Vernam cipher to encrypt communications. Since the bits used to
test for eavesdropping are communicated over the open public
channel, they must always be discarded and only the remaining bits
constitute the key. An intercept-resend attack is not the optimal
eavesdropping strategy. However, any interaction with the data
carriers that can provide Eve with any information on the key
always cause errors in transmission.

In order to leave the original states intact, Eve could try to
attach a probe and let it interact with the information carrier:
\begin{eqnarray}
|a\rangle |E\rangle &\rightarrow& |a\rangle |E_a \rangle \quad
\rm{and} \nonumber
\\|b\rangle |E\rangle &\rightarrow& |b \rangle |E_b \rangle,
\end{eqnarray}
where $|a\rangle$ and $|b\rangle$ denote two possible
states of information carrier, $|E \rangle$ is the initial
state of Eve's probe, and $|E_a \rangle$ and $|E_b \rangle$
are its final states. Any unitary interaction has to
conserve the following inner product
\begin{equation}
\langle a|b \rangle \langle E|E \rangle = \langle a|b \rangle
\langle E_a|E_b \rangle.
\label{in_prod_eq}
\end{equation}
If the states $|a\rangle$ and $|b\rangle$ are
non-orthogonal, $\langle a|b\rangle \neq 0$, the equality
(\ref{in_prod_eq}) can be fulfilled only if $\langle
E_a|E_b \rangle = 1$, i.e., when the final states of Eve's
probe are identical. Eve thus cannot gain any information.
It is apparent that for Eve to discriminate between two
nonorthogonal states she must disturb the state of the
measured objects, and thereby inevitably cause errors in
transmissions. A more detailed discussion of sophisticated
eavesdropping strategies will be provided in Section
\ref{Security}.

It should be mentioned that no physical apparatus is
perfect and noiseless. Alice and Bob will always find
discrepancies, even in the absence of Eve. As they cannot
set apart errors stemming from eavesdropping and those from
the noise of the apparatus, they conservatively attribute
all the errors in transmissions to Eve. From the number of
errors, the amount of information that has potentially
leaked to Eve can be estimated. Afterwards Alice and Bob
reconcile their bit strings using an error correction
technique to arrive at an identical sequence of bits. This
sequence is not completely secret. Eve might have partial
knowledge about it. To eliminate this knowledge, they run a
procedure called privacy amplification. Privacy
amplification is a method enabling them to distill a secret
bit string from their data in such a way that Eve would know
even a single bit of the distilled string only with an
arbitrarily small probability. Both of these procedures,
error correction and privacy amplification, will be
described in detail in Section \ref{sec_procedures}.

\section{Some other discrete protocols for QKD}

\subsection{Two-state protocol, B92}\label{sec_B92}

Besides BB84, other protocols were designed. In 1992, C.\,H.
Bennett showed (\cite{Bennett92}) that two nonorthogonal states
are already sufficient to implement secure QKD. Let Alice choose
two nonorthogonal states and send them to Bob in random order.
When Bob performs projections onto subspaces orthogonal to the
signal states, he sometimes learns Alice's bit with certainty and
sometimes he obtains an inconclusive outcome. After the
transmission, Bob tells Alice when he detected a bit. In this
case, he does not announce the used basis, because a basis in
which he detected a photon, uniquely identifies the bit Alice had
sent. This protocol is usually called B92.

However, such a scheme is secure only in lossless systems or if
the losses are very low. in the case of higher losses, an
eavesdropper could sit in the middle and make measurements on the
quantum states. If she has obtained an inconclusive result, she
blocks the signal, while if she has detected the sent state, she
re-sends a correct copy to Bob, because she knows the state with
certainty. To compensate for the blocked photons, she can send a
pulse of higher intensity so that Bob cannot observe any decrease
in the expected transmission rate.

\subsection{B92 protocol with a strong reference pulse}

One possibility to counteract the above mentioned eavesdropping
strategy against the B92 protocol is to encode bits into a phase
difference between a dim pulse (with less than one photon in
average) and a classical strong reference pulse
(\cite{Bennett92}). It means the laser pulse is split into strong
and weak parts on a highly unbalanced beam splitter. Both Alice
and Bob can introduce a phase shift between these pulses. On Bob's
side both pulses are combined again on an unbalanced beam splitter
where they interfere. Bob can also monitor the presence of all
strong pulses.

Now, when Eve gets an inconclusive result, she cannot
suppress the strong pulse, because Bob must receive all of
them. However, when Eve blocks only the dim pulse,
interference of the bright pulse with vacuum (instead of
the dim pulse) will lead to errors. Similarly, if Eve tries
to fabricate her own dim or bright pulse (or both of them)
and send it (them) to Bob she will inevitably cause
detectable errors. Even though the B92 protocol can be
unconditionally secure if properly implemented, Eve can
acquire now more information on the key for a given disturbance
than in the the case of the BB84 protocol (\cite{fuchs97a}).

\subsection{Six-state protocol}

In the six-state protocol, three non-orthogonal bases are used
(\cite{Bruss98,bechmann99}) that Alice and Bob randomly alternate.
If we denote the two conjugate bases employed in the BB84 protocol
as $\{ \ket{0}, \ket{1} \}$ and $\{ \ket{\bar{0}}, \ket{\bar{1}}
\}$, where
\begin{equation}
\ket{\bar{0}} = \frac{1}{\sqrt{2}} \left( \ket{0} + \ket{1} \right), \quad
\ket{\bar{1}} = \frac{1}{\sqrt{2}} \left( \ket{0} - \ket{1} \right),
\end{equation}
then the third basis is $\{\ket{\bar{\bar{0}}}, \ket{\bar{\bar{1}}} \}$ with
\begin{equation}
\ket{\bar{\bar{0}}} = \frac{1}{\sqrt{2}} \left( \ket{0} + i\ket{1} \right), \quad
\ket{\bar{\bar{1}}} = \frac{1}{\sqrt{2}} \left( \ket{0} - i\ket{1} \right).
\end{equation}
The probability that Alice and Bob choose the same basis is 1/3
now.\footnote{Factors like 1/3 for the six-state protocol or 1/2
for the BB84 are not essential. In fact, the communication can
proceed in only one orthogonal basis and the other non-orthogonal
states can be send randomly from time to time just to test the
channel for the presence of an eavesdropper. So if the
probabilities of bases are ``biased'' in favor of one of the
bases, these factors can asymptotically reach unity
(\cite{efficient}).} But this disadvantage against BB84 is
outweigh by the fact that eavesdropping causes higher error rate.
For example, a continuous intercept-resend attack induces in
average 33\,\% of errors compared to 25\,\% in the case of the
BB84 protocol. In general, the maximal mutual information between
Eve and Alice is smaller than in the BB84 scenario. Besides, the
symmetry of the signal states simplifies the security analysis.

\subsection{SARG protocol}\label{sec_SARG}

The SARG protocol (called after the names of its authors) was
proposed to beat the photon-number splitting attack
(PNS)\footnote{In the photon-number splitting attack Eve exploits
multi-photon states present in weak laser pulses. See
Section~\ref{sec_pns}.} in QKD schemes based on weak laser pulses.
It relies on Eve's inability to perfectly distinguish between two
non-orthogonal states (\cite{SARG04,BGKS05}). In contrast to BB84,
two values of a classical bit are encoded into pairs of
non-orthogonal states. However, to implement the SARG protocol one
can keep the same hardware as for BB84 and modify only the
classical communication between Alice and Bob. Alice prepares four
quantum states and Bob makes measurements exactly as in the BB84
protocol. But Alice does not reveal the basis but the pair of
non-orthogonal signal states such that one of these states is the
one she has sent. Bob guesses correctly the bit if he finds a
state orthogonal to one of two announced non-orthogonal states
(for details see \cite{SARG04}). In comparison with the BB84
protocol, SARG enables to increase the secure QKD radius when the
source is not a single-photon source.

\subsection{Decoy-state protocols}\label{sec_decoy}

The decoy-state method represents another way for counteract the
PNS attack on QKD schemes using weak laser pulses
(\cite{H03,wang04suba,W04,LMC05,M05}). It can substantially
prolong the distance to which the secure communication is
possible. If this method is used with the BB84 protocol the
secure-key rate is proportional to the overall transmittance even
if the light source is an attenuated laser (the secure-key rate
for standard BB84 is linearly dependent on transmittance only in
the case of single-photon source, with weak laser pules it is
proportional to the \emph{square} of the transmittance).

The idea is based on the observation that by adding some decoy
states, one can estimate the behavior of vacuum, single-photon,
and multi-photon states individually. Hence, Alice sends sometimes
an additional, decoy, state with a different intensity than the
states used for the key transmission (but with the same
wavelength, timing, etc.). These decoy states serve only for
testing Eve's presence. Eve does not know when Alice sends the
decoy states and she cannot identify them. Changes, that Eve's PNS
attack makes on these decoy states, enable Alice and Bob to detect
the PNS eavesdropping.

The essence of the decoy-state method consists in the following
fact: The conditional probability $Y_n$ that Bob detects a signal
-- providing that Alice's source has emitted an $n$-photon state
-- must be the same both for the signal and decoy states. When no
eavesdropper is present it must be equal to the following value
given by the parameters of the apparatus:
\begin{equation}
Y_n^{\mathrm{signal}} = Y_n^{\mathrm{decoy}} = Y_n = [1 -
(1-\eta)^n] (1-p_{\mathrm{dark}}) + p_{\mathrm{dark}},
\end{equation}
where $\eta$ is the total transmission efficiency and
$p_{\mathrm{dark}}$ is the probability of the detector dark count.
The PNS attack inevitably changes some $Y_n$. The quantities $Y_n$
are not directly measurable. But what Bob can directly determine
is the total detection rate for a given mean photon number $\mu$
of Alice's pulses:
\begin{equation}
  Q_\mu = e^{-\mu} \sum_{n=0}^\infty Y_n {\mu^n \over n!}.
\end{equation}
If Alice and Bob use decoy states with different mean photon
numbers they can estimate values of $Y_n$ for some photon numbers
$n$ and check whether they correspond to the expected values.

The security of the decoy-state method with the BB84 protocol
under the ``paranoid'' assumptions (\cite{gottesman04a}) has been
analyzed by \cite{LMC05}.

\subsection{Entanglement-based protocols}\label{EntPairs}

Another class of QKD protocols is based on quantum entanglement.
The security of the original proposal was ensured by checking the
violation of Bell's inequalities (\cite{Ekert91}). The simplified
version of the protocol works in a very similar way as BB84
(\cite{EPR-BB84}).

\subsubsection{Entanglement, Bell's inequalities}

Two or more quantum systems are entangled if their global
state cannot be expressed as a direct product or a
statistical mixture of direct products of any quantum
states of individual systems. Entanglement leads to many
interesting effects unknown in classical physics. It lies
in the basis of quantum teleportation (\cite{teleport}) and
it is responsible for the effectiveness of quantum
computation (\cite{Nielsen}). Asher Peres said that
``Entanglement is a trick that quantum magicians use to
produce phenomena that cannot be imitated by classical
magicians.'' (\cite{Bruss2002}).

In 1935 Einstein, Podolsky and Rosen (\cite{EPR})
formulated a gedanken experiment employing two particles
prepared in an entangled state to argument against the
completeness of quantum theory. They used the fact that the
result of any potential measurement on one subsystem of the
properly chosen entangled pair can be predicted with
certainty after the proper measurement on the other
subsystem. Following this fact and a few ``natural''
assumptions (namely the assumptions of locality and
reality) they concluded that there must simultaneously
exist ``elements of reality'' for two \emph{complementary}
observables.

However, in 1964 John Bell (\cite{Bell}) has shown that
there is no local realistic theory that would give the same
predictions as quantum mechanics. Namely, quantum mechanics
predicts different values of certain correlations of
measurement results on a bipartite system in a specific
entangled state. He derived his famous inequalities that
must be satisfied by any local realistic theory but that
may be violated by quantum theory.

Let us denote ${\sl A}({\mathbf{n}}_1)$ and ${\sl
B}({\mathbf{n}}_2)$ random variables, getting discrete
values $\pm 1$, corresponding to measurement results on two
separated but somehow correlated particles, where the
settings of respective measurement devices are represented
by unit vectors ${\mathbf{n}}_1$ and ${\mathbf{n}}_2$ (note
that ${\sl A}$ depends only on ${\mathbf{n}}_1$ and ${\sl
B}$ only on ${\mathbf{n}}_2$ -- this reflects the locality
condition). The randomness of ${\sl A}$ and ${\sl B}$ is
supposed to be caused only by some random parameters
$\lambda$ that may be common for both the particles and
that we do not know (the premise of reality). The Bell
inequality, in the form derived by \cite{CHSH}, states that:
\begin{equation}
   | C({\mathbf{n}}_1, {\mathbf{n}}_2) + C({\mathbf{n}}'_1, {\mathbf{n}}_2)  +
     C({\mathbf{n}}_1, {\mathbf{n}}'_2) - C({\mathbf{n}}'_1, {\mathbf{n}}'_2) | \le
     2,
 \label{bell_ineq}
\end{equation}
where $C({\mathbf{n}}_{1},{\mathbf{n}}_{2})$ is the
correlation function:
\begin{equation}
  C({\mathbf{n}}_{1},{\mathbf{n}}_{2}) = \langle {\sl A}({\mathbf{n}}_1)
    {\sl B}({\mathbf{n}}_2) \rangle
  = \int A({\mathbf{n}}_{1},\lambda)
    B({\mathbf{n}}_{2},\lambda) \, {\mathrm{d}} \rho_\lambda.
 \label{bell_corr}
\end{equation}

Now, let us try to describe such a situation by the quantum
language, assuming two spin-half particles in the following
entangled state:
\begin{equation}
    \ket{\psi} = {1 \over \sqrt{2}} \Bigl( \ket{{\mathbf{n}},+}_1
    \ket{{\mathbf{n}},-}_2 - \ket{{\mathbf{n}},-}_1 \ket{{\mathbf{n}},+}_2
    \Bigr),
 \label{bell_state}
\end{equation}
where state vectors $\ket{{\mathbf{n}},\pm}$ correspond to
two orthogonal projections of spin to direction
$\mathbf{n}$. Then the quantum prediction for correlation
function reads:
\begin{equation}
  C({\mathbf{n}}_{1},{\mathbf{n}}_{2}) = \langle \psi | ({\mathbf{n}}_1
  \cdot \bm{\sigma}_1) ({\mathbf{n}}_2 \cdot
  \bm{\sigma}_2) \ket{\psi},
 \label{bell_qm}
\end{equation}
where $\bm{\sigma}_1, \bm{\sigma}_2$ are vectors of Pauli
matrices. If we choose the settings of the measurement
apparatuses in such a way that ${\mathbf{n}}_2$ with
${\mathbf{n}}_1$, ${\mathbf{n}}_1$ with ${\mathbf{n}}'_2$
and ${\mathbf{n}}'_1$ with ${\mathbf{n}}_2$ include angle
$45^\circ$, while ${\mathbf{n}}'_1$ with ${\mathbf{n}}'_2$
include angle $135^\circ$, we readily find that
 \begin{equation}
   | C({\mathbf{n}}_1, {\mathbf{n}}_2) + C({\mathbf{n}}'_1, {\mathbf{n}}_2)  +
     C({\mathbf{n}}_1, {\mathbf{n}}'_2) - C({\mathbf{n}}'_1, {\mathbf{n}}'_2) |
    = 2 \sqrt{2} > 2.
 \label{Bell_viol}
 \end{equation}

\subsubsection[Ekert's protocol and its
      simplified form]{Original Ekert's protocol and its
      simplified form}

According to Ekert's protocol (\cite{Ekert91}), Alice and Bob each
obtain one particle from a pair of spin-1/2 particles in the state
(\ref{bell_state}). (In fact, it does not matter whether they
share two entangled spin-1/2 particles or, e.g., two photons with
entangled polarizations.) Alice and Bob perform measurements on
their respective particles in three bases defined by three
orientations of their measurement devices (e.g., Stern-Gerlach
apparatuses). For simplicity let us suppose that they use only
directions lying in the plane perpendicular to the trajectory of
the particles. Alice's bases make angles with respect to the
vertical $0^\circ, 45^\circ, 90^\circ$, and Bob's bases are making
$45^\circ, 90^\circ, 135^\circ$. There are nine possible
combinations. After the quantum transmission, during which Alice
and Bob randomly and independently set their measurement bases,
the settings are publicly announced. When identical bases were
used, the outcomes of their measurements are correlated and become
the cryptographic key. The probability that Alice and Bob use the
same basis is 2/9. The outcomes of measurements in the other bases
are used to verify the violation of the Clauser-Horne-Shimony-Holt
inequality (\ref{bell_ineq}). An eavesdropper attempting to
correlate his probe with the other two particles would disturb the
purity of the singlet state (\ref{bell_state}), which would result
in a smaller violation of the inequality or no violation at all.

A year later \cite{EPR-BB84} proposed a simpler entanglement-based
protocol without invoking directly Bell's theorem. Here, both
Alice and Bob choose only from two bases corresponding to two
perpendicular orientations of their spin-measurement devices in a
way very similar to BB84 protocol. In fact, the only difference
from BB84 is that Alice does not \emph{send} particles in a chosen
spin (or polarization) state but she \emph{measures} her particle
from the entangled pair in one of two conjugated bases. She must
select bases randomly and independently from Bob. The rest is the
same as in BB84: After the transmission Alice and Bob compare
their bases and keeps only those results when they used the same
bases.

\subsubsection{Passive setup}\label{sec_passive}

The system for entanglement-based QKD can be designed even
in such a way that it can be operated entirely in a passive
regime without any extern-driven elements (e.g.,
polarization rotators or phase modulators; \cite{passive}).
Each particle from the entangled pair ``may freely decide''
on a beamsplitter in which basis it will be measured. It
means, both the random key bits and random measurement
basis are chosen directly by the genuine randomness of the
nature.

\section{Experiments}\label{sec_experiments}

\subsection{QKD with weak laser pulses}

Attenuated lasers are often used as sources in practical QKD
devices. If the spectral width of the laser pulses is much smaller
than their mean frequency, the state of light can well be
approximated by a monochromatic coherent state. The photon-number
distribution of the coherent state is governed by the Poisson
statistics. The multi-photon pulses can cause problems due to the
PNS attack. Eve could always split off one photon and perform a
measurement on it without introducing an error. This potentially
leaked information must be taken into account (see
Sections~\ref{sec_BS} and \ref{sec_pns}). The trick how to beat
this attack appears in the decoy-state method (see
Section~\ref{sec_decoy}).

\subsubsection{Polarization encoding}

The very first QKD experiment that took place in 1989
(\cite{BB-exp,ErrCorr1}) was based on polarization encoding for the BB84 protocol. For
the description of the protocol, we refer the reader to Section
\ref{QKD}.

\begin{figure}
\centerline{\resizebox{0.9\hsize}{!}{\includegraphics{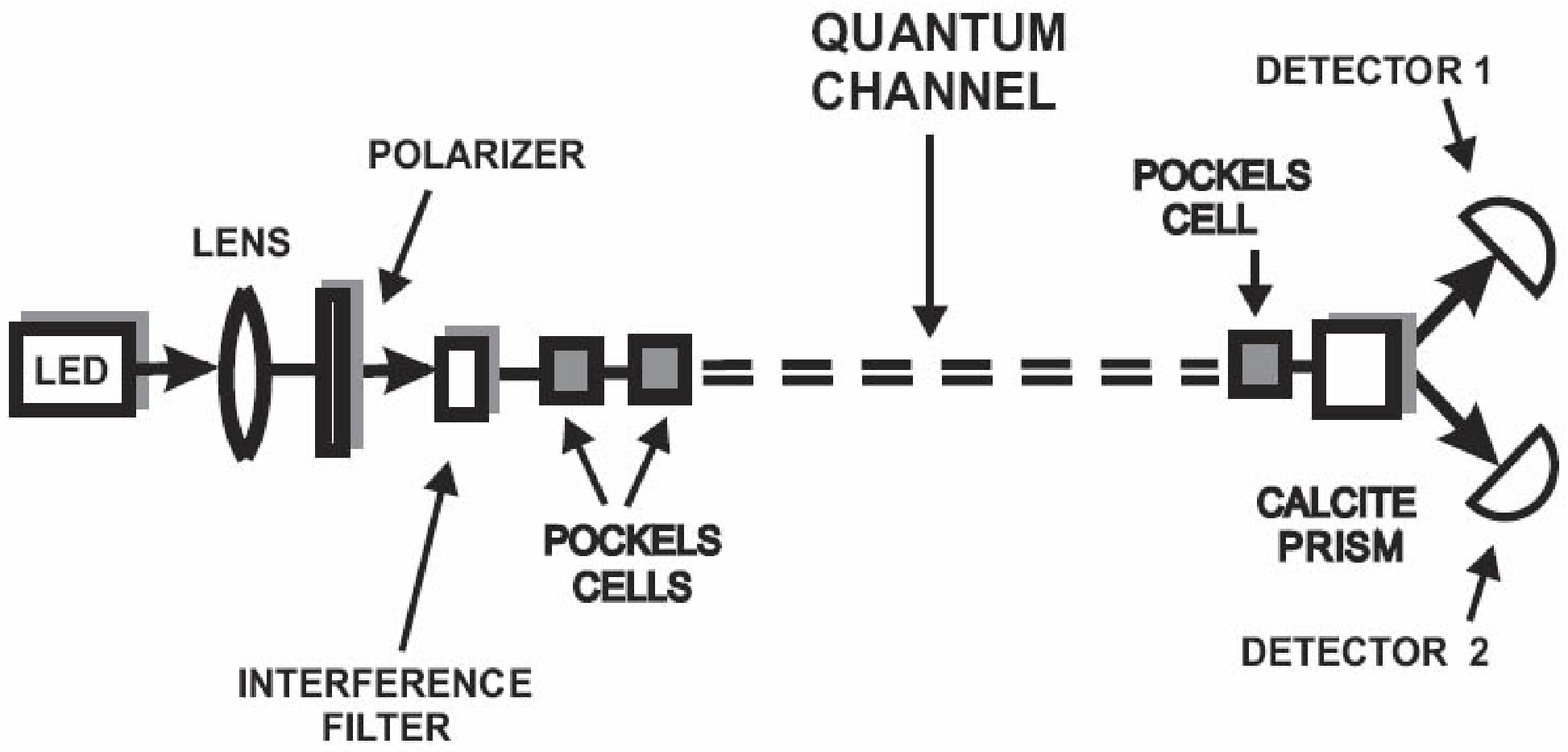}}}
\caption{First QKD experiment (\protect\cite{BB-exp}).}
\label{Fig_FirstQKD}
\end{figure}

A  light-emitting diode (LED) generated light pulses that were subsequently
attenuated by an interference filter and polarized by a
polarizer (see Fig.~\ref{Fig_FirstQKD}). The qubits were
encoded in the polarization of photons by means of Pockels
cells. The quantum channel was 32\,cm of free air. Bob
analyzed the polarization states using a Wollaston prism,
which was preceded by another Pockels cell to choose his
polarization basis. The output ports of the prism were
monitored by photomultipliers.

Four years later, Gisin's group from the University of
Geneva replaced the free-air optical path by a 1\,km
optical fiber (\cite{Muller93,Breguet94}). A semiconductor
laser at 800\,nm was used to generate light pulses that
were detected by silicon avalanche photodiodes. Since the
optical fiber deforms the polarization state of light, a
manually adjustable polarization controller was employed to
compensate for temporal changes of polarization.

Bends and twists of the optical fiber induce birefringence,
which gives rise to different velocities of the orthogonal
polarization components of light that result in the change
of the polarization state. Since the degree of polarization
degrades slowly in fibers, the same stress-induced
birefringence can, on the other hand, be used to compensate
for this deformation. A fiber spool of a suitable diameter
can act as a fractional wave plate.

\cite{Franson94} proposed a QKD device with an active
polarization-alignment feedback loop. Such a system was
demonstrated to work over a distance of 1\,km (\cite{Franson95}).

The first experiment with  Alice and Bob being placed in different
laboratories (in this case even different towns of Geneva and
Nyon) was performed by the Geneva group
(\cite{Muller95,Muller96}). Error rates of only 3--4 \% were
achieved between two stations, connected by a 23\,km fiber
deployed under Lake Geneva. In order to reduce fiber losses, a
laser at 1.3\,$\mu$m was used and the photons were detected by
liquid-nitrogen-cooled germanium avalanche photodiodes.

Using optical fiber is not the only way to implement QKD at a
distance. Another approach is to try to communicate directly
through free space. Unlike fibers, the atmosphere is
non-birefringent, thereby polarization encoding is very suitable.
The feasibility of free-space QKD was shown by \cite{Jacobs96},
who managed to communicate over 150\,m in a
fluorescent-tube-illuminated corridor and over 75\,m outdoors in
daylight. It was the first free-space implementation of QKD after
the celebrated 1989 Bennett and Brassard experiment and there were
more to come. The Los Alamos group first exchanged keys at 1\,km
by night bouncing the photons between mirrors
(\cite{Buttler98a,Buttler98b}), then point-to-point communication
over 0.5\,km in daylight was performed (\cite{Hughes00}) and
eventually over 1.6\,km in daylight (\cite{Buttler00}). The
distance 1.9\,km at night were covered by \cite{Gorman01}.
\cite{hughes02} then demonstrated free-space QKD over 10\,km.
Free-space QKD over the largest distance so far was performed by
the Munich group of H.~Weinfurter (\cite{Munich,munich-nature}).
Unlike the other groups, they moved to the high altitudes of the
Alps to take advantage of thinner air and less air turbulence.
Alice was located on the summit of Zugspitze (2962\,m) and Bob was
on a 23.4\,km distant Karwendelspitze (2244\,m).

Demonstration of free-space QKD with a single-photon source based
on a nitrogen-vacancy center in diamond (see
Section~\ref{sec_color_centers}) was done by \cite{centers_QKD1}
(indoor experiment over 50\,m) and by \cite{centers_QKD2} (this
later experiment of the same group was operated outdoors over
30\,m at night).

\subsubsection{Phase encoding}

In this method, different polarizations (used in polarization
encoding) are replaced by different phase shifts between two arms
of the Mach-Zehnder interferometer. Alice controls the phase shift
in one arm of the interferometer, Bob controls the phase shift in
the other arm. If Alice's and Bob's phase shifts are the same or
differ by $180^\circ$, then the behavior of the photon at Bob's
beam splitter is deterministic because of constructive
interference in one of the outputs and destructive interference in
the other one. If the total phase shift between the arms is
different from an integer multiple of $180^\circ$,  photons are
detected randomly at both detectors.

in the case of the BB84 protocol, Alice encodes bit values into
four non-orthogonal quantum states. She sends weak light pulses to
the interferometer and sets randomly phase $\phi_A$ to $0^\circ$,
$90^\circ$, $180^\circ$, or $270^\circ$. Bobs sets randomly (and
independently of Alice) phase $\phi_B$ to $0^\circ$ or $90^\circ$.
These two values correspond to the measurement in ``rectilinear''
and ``diagonal'' bases, respectively:
\begin{center}
\begin{tabular}{|lccc|ccr|}
  \hline
   $\phi_A$~ & \RB:~ &  $~0^\circ$ \dots ``1'', & $180^\circ$ \dots  ``0'' &
   $\phi_B$~ & \RB:~ &  $0^\circ$ \\
              & \DB:~ & $90^\circ$ \dots ``1'', &  $270^\circ$ \dots ``0'' &
              & \DB:~ & $90^\circ$ \\
  \hline
\end{tabular}
\end{center}

\begin{figure}
\centerline{\resizebox{\hsize}{!}{\includegraphics{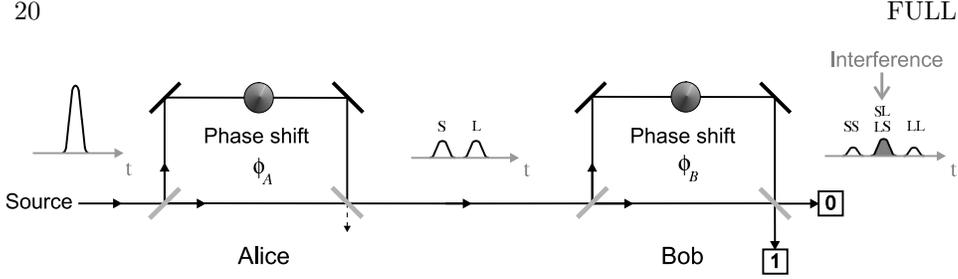}}}
\caption{Setup for phase-encoded QKD with a double
Mach-Zehner interferometer.} \label{fig_double_MZ}
\end{figure}

However, in practice it is impossible to keep the same and stable
phase conditions in two different arms of the Mach-Zehnder
interferometer over long distances. The way how to solve this
problem was proposed already by \cite{Bennett92}. Two
communicating parties can employ a time multiplex and use only one
optical fibre to interconnect their devices (see
Fig.~\ref{fig_double_MZ}). Now two unbalanced Mach-Zehnder
interferometers are used. The path lengths' difference between the
longer and shorter arm of each interferometer is larger than the
width of the laser pulse.\footnote{If the pulse width is in the
order of nanoseconds then the path lengths' difference is usually
a few meters.} But the path differences are the same for both
interferometers. The case where the photon goes first through the
longer (L) arm and then through the shorter (S) one is
indistinguishable from the case when it first passes the shorter
and then the longer arm. This path indistinguishability results in
the interference at the last beam splitter. Thus for the ``central
peak'' (see the right side of Fig.~\ref{fig_double_MZ}) the system
behaves exactly in the same way as a single balanced Mach-Zehnder
interferometer. This peak is selected by the proper timing of
detection and the events when the photon passed either through
both shorter or both longer arms are ignored.

The first system based on phase encoding was build by
\cite{townsend93a} (see also \cite{townsend93b}). The signal was
sent through 10\,km of fiber in a spool. Later the system was
modified so that the polarization in long arms was rotated by
$90^\circ$ in both interferometers and the time multiplex was
supplemented by a polarization multiplex. That is, at the output
of Alice's interferometer and at the input of Bob's interferometer
there were polarization beam splitters. This technique suppresses
the lateral non-interfering peaks (\cite{townsend94}). Further the
distance were prolonged to 30\,km (\cite{marand95}).
\cite{townsend97} also tested a wavelength-division multiplex to
execute both the QKD and the classical communication through the
same fiber on different wavelengths. A QKD system with a double
Mach-Zehnder interferometer was realized also in Los Alamos
National Laboratory (\cite{hughes96,hughes00b}). They tested it in
an installed optical fiber up to a distance of 48\,km. Another
fiber-based system (at 830\,nm) was realized by \cite{Dusek}. It
had implemented an active stabilization of interferometers and
programmed all supporting procedures for practical QKD. The system
was used as a quantum identification system (for mutual
identification of the users) at a distance of 500\,m. The system
with silica-based integrated-optic interferometers was built by
\cite{kimura} and tested at a distance over 150\,km. Toshiba
Research Europe developed an automated system at 1550\,nm with a
new method for active interferometer stabilization (a
``stabilization'' pulse goes after each signal pulse) and tested
it at the distances up to 122\,km (\cite{gobby04,yuan05}).

The systems using either the polarization encoding or double
Mach-Zehnder interferometer require an \emph{active} stabilization
to compensate drifts and fluctuations of polarizations and/or
phases. \cite{muller97} has proposed an interesting way how to
implement QKD device (using a phase encoding) where all optical
and mechanical fluctuations are automatically \emph{passively}
compensated (the principle of this auto-compensation is based on
an earlier idea of \cite{Martinelli}). Two strong mutually delayed
pulses of orthogonal linear polarizations go from Bob to Alice. At
Alice's side they are attenuated (a part of them is also used to
synchronization purposes), the first pulse is phase shifted (this
is the way Alice encodes the information), and both pulses are
reflected on a Faraday mirror. The Faraday mirror, which is a
Faraday rotator followed by a mirror, exchanges their vertical and
horizontal polarization components. Then these two dim pulses
return to Bob. Because they go back through the same line but have
properly modified polarizations by the Faraday mirror, all the
polarization distortions caused by birefrigence experienced by the
pulses in their first trip are compensated during the return trip.
At the end the sent vertical polarization returns as horizontal
one and vice versa. At Bob's side the first pulse passes a longer
arm of an unbalanced Mach-Zehnder interferometer while the second
pulse passes its shorter arm (the pulses are separated by a
polarization beam splitter and then their polarizations are made
the same). In one of the arms Bob now applies his phase shift.
Because the original delay between the pulses was created by the
same unbalanced interferometer no stabilization of this
interferometer is needed. Since no special optical adjustment is
necessary to operate this set-up it is usually called
``plug\&play'' system. However, there are also some drawbacks: The
fact that pulses must go first from Bob to Alice and then back
complicates the timing of the whole process and may effectively
decrease the transmission rate. The problem is, especially, with a
Rayleigh backscattering. To suppress its contribution to error
rate the strong pulses coming form Bob should not meet with the
weak pulses propagating in the opposite direction. Further,
because the strong pulses must pass the whole path from Bob to
Alice before they are attenuated and the information is encoded,
Eve has an opportunity to change some of their properties, e.g.,
their photon statistics. The system is also more sensitive to a
certain ``Trojan horse'' attacks (see Section~\ref{sec_side_ch}).

The first experimental realization was done by \cite{zbinden97}.
The key was exchanged over a 23-km-long optical fiber installed
under Lake Geneva. Later the fully automated system was tested on
the same fiber (\cite{ribordy00}). The implemented protocol was
BB84. The system was operated at 1300\,nm. A similar
auto-compensating system operating at 1300\,nm was also
independently developed at IBM (\cite{bethune00}). It was tested
on a 10-km-long fiber in a spool. In this set-up the pulses sent
by Bob had a reduced intensity to avoid Rayleigh backscattering.
Synchronization was provided by classical pulses at 1550\,nm using
a wavelength-division multiplex. \cite{nielsen01} built a system
working at 1310\,nm and distributed a key over 20\,km in fiber.
Group of A. Karlsson demonstrated that the plug\&play technique
can be implemented in fibers also at 1550\,nm
(\cite{bourennane99}). Later the operation of an improved Geneva
plug\&play setup at 1550\,nm was demonstrated over a 67-km-long
optical-fiber link between Geneva and Lausanne (\cite{stucki02}).

The first experimental demonstration of the decoy-state method
(see Section~\ref{sec_decoy}) was done by \cite{decoy-exp}. Their
set-up used a modified commercial QKD ``plug\&play'' system
manufactured by id Quantique. The distribution was tested over the
distance of 15\,km. The protocol was based on the BB84 scheme
together with a practical implementation of the decoy-state method
with only one decoy state. The average intensities of the signal
and decoy states were chosen to be 0.8 and 0.12 photons,
respectively. Roughly 88\,\% of signal states and 12\,\% of decoy
states were transmitted.

\cite{gisin04} proposed a new technique for practical QKD, based
on a specific protocol and tailored for an implementation with
weak laser pulses. The key is obtained by a simple measurement of
the times of arrival of the pulses incoming to Bob. The presence
of an eavesdropper is checked by an interferometer built on an
additional monitoring line. Each logical bit is encoded into a
sequence of two pulses: either one empty and one non-empty or vice
versa. There is a phase coherence between any two non-empty pulses
because a mode-locked laser is used as a source. Some pulses are
reflected at Bob's beam-splitter and go to the unbalanced
Mach-Zehnder interferometer (monitoring line). Here is where
quantum coherence plays a role. If coherence is not broken, only
the detector at the particular output of the interferometer may
fire at certain instants. This enables to detect an eavesdropping.
The first experimental realization of this protocol was done by
\cite{stucki05}.

\subsection{Entanglement-based protocols}

The principle of entanglement-based protocols was explained in
Section~\ref{EntPairs}. In practical realizations only the
entangled states of photons are used. However, different kinds of
entanglement can be employed: For example, entanglement in
polarizations of photons, entanglement in energy and time,
entanglement in orbital angular momentum, or so called
``time-bin'' entanglement which is a special case of energy-time
entanglement. Experiments with QKD using photon pairs often
utilized set-ups and took up on experiments examining the
violation of Bell's inequalities. Besides QKD, the distribution of
entanglement between distant users can be beneficial also for
other task like quantum teleportation, quantum dense coding,
quantum secret sharing, etc. However, there is a problem of
coupling between the property used to encode the qubits and the
other properties of the carrier electromagnetic field, that rises
during the propagation in a dispersive medium. This form of
decoherence gradually destroys quantum correlations between the
photons.\footnote{This effect has also a positive aspect: It
prevents unintentional information leakage in unused degrees of
freedom (\cite{mayers98}).} For example, polarization-mode
dispersion makes two values of polarization-encoded qubit
distinguishable also in temporal domain and so wipes out quantum
correlations between polarizations. Similarly, chromatic
dispersion degrades energy-time entanglement.

\subsubsection{Polarization entanglement}

In this case Alice and Bob are each provided by one photon of an
entangled pair of one of these forms:
\begin{equation}
  \frac{1}{\sqrt{2}} \left( \ket{V}_{\mathrm{A}}
    \ket{V}_{\mathrm{B}}
    \pm \ket{H}_{\mathrm{A}} \ket{H}_{\mathrm{B}} \right), \quad
  \frac{1}{\sqrt{2}} \left( \ket{V}_{\mathrm{A}}
    \ket{H}_{\mathrm{B}}
    \pm \ket{H}_{\mathrm{A}} \ket{V}_{\mathrm{B}} \right),
\end{equation}
where $\ket{V}, \ket{H}$ denotes single-photon states with
vertical and horizontal linear polarizations, respectively. The
pairs are prepared by a parametric down-conversion process in
nonlinear optical crystals. Polarization entanglement is created
either by one crystal using the phase matching of type-II (in a
proper geometrical lay-out) or by two crystals with type-I phase
matching that are placed closely one by one but with optic axes
oriented perpendicularly to each other. Alice and Bob are equipped
with polarization analyzers that can rapidly change measurement
polarization bases, e.g., electro-optical polarization modulators
followed by polarizing beam splitters (with photon counters behind
them).

The first two experiments were reported in 2000. Zeilinger's group
 (\cite{jennewein00}) used a
BBO\footnote{$\beta$-BaB${}_{2}$O${}_{4}$.} crystal, cut for
type-II phase matching and pumped by an argon-ion laser, to
generate photon pairs at 702\,nm (both photons had the same
wavelength). Their analyzers consisted of fast modulators,
polarizing beam splitters, and silicon avalanche photodiode (APD)
detectors. They have demonstrated QKD over 360\,m in installed
single-mode fibers. Kwiat's group in Los Alamos (\cite{naik00})
worked with two BBO crystals of type-I phase matching  pumped by
an argon-ion laser and they also produced photon pairs with
degenerate wavelengths at 702\,nm. They implemented original Ekert
protocol and have demonstrated QKD in free space at the distance
of a few meters. In addition, they simulated experimentally
different eavesdropping strategies. A newer experiment was done by
\cite{poppe04} in Vienna. Secret key was distributed over
1.45-km-long installed fiber (between a bank and the City Hall).
Polarization-entangled pairs at 810\,nm were produced by type-II
parametric down conversion in a BBO crystal pumped by a
semiconductor laser. The distribution of entanglement over 13\,km
in free space was demonstrated by \cite{peng04}. It was use both
to prove a space-like separated violation of Bell's inequality and
to realize QKD based BB84-like protocol. It utilized type-II
parametric down-conversion in BBO crystal pumped by an argon-ion
laser. Wavelengths of entangled photons were 702\,nm.

\subsubsection{Energy-time entanglement, phase encoding}

Now the employed two-photon entangled states have the approximate
form:
\begin{equation}
 \int \mathrm{d} \omega \, \xi(\omega) \, \ket{\omega}_{\mathrm{A}}
   \ket{\omega_0-\omega}_{\mathrm{B}},
\label{e-t_ent}
\end{equation}
where $\ket{\omega}$ denotes a single-photon state at frequency
$\omega$, $\omega_0$ is an optical frequency of the pump laser,
and $\xi(\omega)$ expresses the distribution of individual
frequency components. The pairs are again produced by parametric
down conversion in nonlinear optical crystals. Photons in states
close to that given by Eq.~(\ref{e-t_ent}) -- neglecting vacuum
and multi-pair contributions -- are generated when the crystal is
pumped by a laser with a large coherence time. Alice and Bob
obtain one photon each and they let them pass through identically
unbalanced Mach-Zehnder interferometers (one interferometer at
Bob's side, one at Alice's side). The path lengths' difference
between the longer and shorter arm of each interferometer must be
larger than the coherence length of generated photons but shorter
then the coherence length of the pump laser. The path differences
must be the same for both interferometers. The instants of
detections of two photons from a pair are very tightly correlated
(of the order of hundreds of femtoseconds) but the particular
times of these coincident detections are uncertain and random.
Therefore Alice and Bob cannot distinguish between the situations
when both photons went through longer arms of their
interferometers and when both of them went through shorter arms
(this leads to the fourth-order interference). Alice and Bob
chooses their measurement bases by changing the phase shifts
between the arms of their interferometers (e.g., they can randomly
and independently alternate shifts $0^\circ$ and $90^\circ$). When
their phase difference is $0^\circ$, the measurement outcomes are
deterministic. When the phase difference is $\pm 90^\circ$, the
results are random. Events when one photon went through a shorter
arm and the other one through a longer arm, are ignored. This
arrangement was originally devised by \cite{franson89} for another
purposes. Its use for practical QKD in fibers was proposed by
\cite{ekert92}. The set-up can be further modified to be operated
completely in a passive way (\cite{passive}) -- see
Section~\ref{sec_passive}.

This QKD scheme was first realized by \cite{ribordy01} from the
University of Geneva. They used a KNbO${}_3$ crystal pumped by a
doubled Nd-YAG laser to create entangled pairs with asymmetric
wavelengths 810\,nm and 1550\,nm. The wavelength 810\,nm gave an
advantage to use efficient and low-noise Si-APD photon counters at
Alice's side (the distance between the source and Alice's analyzer
was very short). The wavelength 1550\,nm of the other photon fit
to low-loss window of optical fibers, so this photon travelled the
longer distance between the source and Bob.  Bob was connected to
the source by 8.5-km-long optical fiber in a spool (the
dispersion-shifted fiber was used to limit the decoherence induced
by chromatic dispersion). It should be noted that the passive
set-up was implemented. Two measurement bases (at each terminal)
were passively randomly selected using a polarizing beam splitter.
One physical interferometer behaved like two interferometers with
different phase settings for two different polarizations of light.

\subsubsection{Time-bin entanglement, phase-time encoding}

\begin{figure}
\centerline{\resizebox{0.8\hsize}{!}{\includegraphics{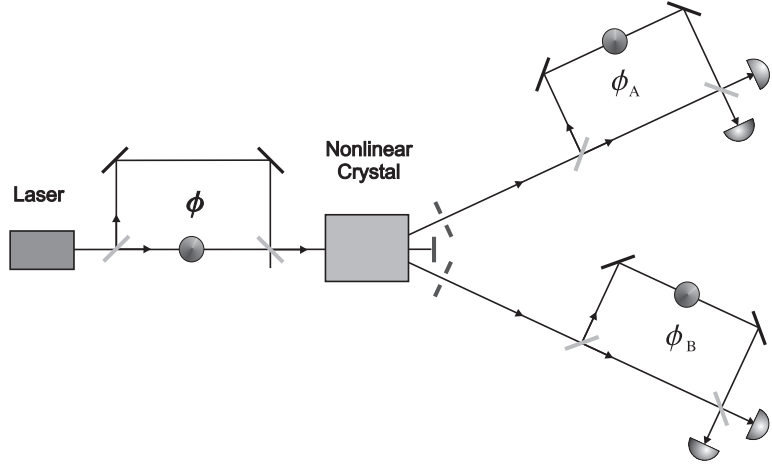}}}
\caption{Schematic setup for QKD using time-bin
entanglement.} \label{fig_time-bin}
\end{figure}

This method is similar to the phase encoding described above. But
now there is one more unbalanced Mach-Zehnder interferometer
placed in the pump beam and a \emph{pulsed} source is used to pump
the crystal. The scheme of the apparatus is shown in
Fig.~\ref{fig_time-bin}. The generated pair can be described by
the following state:
\begin{equation}
    \frac{1}{\sqrt{2}} \left( e^{i\phi} \, \ket{S}_{\mathrm{A}}
    \ket{S}_{\mathrm{B}}
    - \ket{L}_{\mathrm{A}} \ket{L}_{\mathrm{B}} \right),
\end{equation}
with $S$ and $L$ denoting contributions from pump pulses going
through a shorter and longer arm of the interferometer,
respectively. The path differences of all three interferometers
should be the same. Now Alice can detect a photon in three
different time windows (after each laser pulse): The first
corresponds to the situation when both the pump pulse and Alice's
photon went through the shorter arms ($SS$), the second
corresponds to the combination of the shorter and the longer arm
or vice versa ($SL$ or $LS$), and the third corresponds to the
situation when both the pump pulse and Alice's photon went through
the longer arms ($LL$). The same holds for Bob's detections. To
establish the secret key Alice and Bob publicly agree on the
events when both of them detected a photon (does not matter at
which detector) either in the first or in the third time window,
but do not reveal in which one, and on the events when they both
registered detector clicks in the second time window, without
revealing at which detector. In the first case they assign
different bit values to the first and third time window (Alice and
Bob must have correlated detection times). The second case (both
photons detected in the second time window) is formally equivalent
to the above described phase-encoding method.

This technique was proposed by \cite{brendel99} (who have also
built the source of pairs) and the QKD experiment was performed by
\cite{tittel00}. The system was tested only in the laboratory. The
crystal KNbO${}_3$ was pumped by a pulsed semiconductor laser
diode. The wavelength of down-converted photons was 1310\,nm.
Later, the distribution of time-bin entangled qubits was
demonstrated over 50\,km of optical fiber (\cite{marcikic04}).

\section{Technology}

\subsection{Light sources}

\subsubsection{Attenuated lasers}

In practical QKD systems the attenuated lasers are still the only
reasonable light sources (except systems using entangled pairs).
The radiation from a laser can be usually well described by a
single-mode coherent state exhibiting Poissonian photon-number
distribution (with $\mu$ being a mean photon number):
\begin{equation}
p(n) = \frac{\mu^n}{n!} e^{-\mu}.
\end{equation}
Clearly, a highly attenuated laser pulse with very small $\mu$
represents a good approximation of a single-photon Fock state (or
rather a superposition of states $\ket{0}$ and $\ket{1}$) because
the ratio $p_{\mathrm{multi}}/p(1)$ of the probability of more
than one photon, $p_{\mathrm{multi}}=\sum_{n=2}^{\infty} p(n)$,
and a single-photon probability, $p(1)$, goes to 0 as $\mu \to 0$.
The only problem is the increasing fraction of vacuum states
($n=0$). For example, if $\mu=0.1$ then $p(0) \doteq 0.905$, $p(1)
\doteq 0.090$, and $p_{\mathrm{multi}} \doteq 0.005$. Empty pulses
decrease transmission rate. A more important problem arises from
detector dark counts. Because detectors must be active for all
pulses including empty ones the dark-count rate is constant while
the rate of non-empty pulses decreases with decreasing $\mu$. This
prevents the use of arbitrarily low mean photon numbers.

The mean photon number must be chosen according to several
aspects. The existence of detector dark counts and the
losses in the system admonish us to use the mean photon
number as high as possible. On the other hand the potential
leakage of information trough the multi-photon pulses
forces us to use the mean photon number as low as possible.
The optimal mean photon number is such that maximizes the
secure-key rate for given conditions. It results from the
trade-off between the value of the detection rate and the
shortening of the key due to privacy amplification (because
of multi-photon contributions, privacy amplification
shortens the resulting distilled key substantially if $\mu$
is too high, namely if $\mu \gtrsim \eta$ where $\eta$ is
the line transmittance; \cite{Lutkenhaus}).

A good measure of the quality of imperfect single-photon sources is the
second-order autocorrelation function of the source, $g_2 =
{\langle I^2 \rangle}/{\langle I \rangle^2}$, i.e., the
correlation measured in a Hanbury-Brown-Twiss-type experiment ($I$
means optical intensity). It can be approximately calculated as
$g_2 \approx 2\,p(2)/[p(1)]^2$ if $p(1) \gg p(2) \gg
\sum_{n=3}^{\infty} p(n)$. The value $g_2 = 1$ corresponds to
Poissonian case, $g_2 < 1$ indicates sub-Poissonian distribution.

\subsubsection[Single-photon sources: PDC]{Single-photon sources: Parametric down conversion}

Another way how to prepare quasi-single-photon states is to use
photon pairs generated by spontaneous parametric down conversion
(SPDC) (\cite{Mandel}). Here the crucial point is a tight time
correlation between photons in the pair. In the ideal case, if one
places a photon-number detector into the path of one member of the
pair (say, into the idler beam) and detects one photon then in the
same time (i.e., in a very short time window of the order of
hundreds of femtoseconds) there must be one photon also in the
other -- signal -- beam.

In reality, due to losses in the signal beam, caused mainly by
an inefficient coupling into the fiber, and partly also due to
dark counts of the trigger detector, there may be no photon in the
signal beam even if the trigger detector has clicked. However, the
probability of this event is relatively low -- today typically
about 30\,\%.

Nearly all practically applicable detectors cannot
distinguish the number of photons and their quantum
efficiency is substantially lower than 100\,\%. Therefore,
there is also non-zero probability having more than one
photon in the signal beam after the trigger detection.
(Notice that the number of photons in one mode is thermally
distributed and the total number in all modes obeys the
Poissonian distribution.) On the other hand, the efficiency
of the conversion of a pump photon into the pair of
sub-frequency photons is very low, typically about
$10^{-10}$, so the probability of generation of
multi-photon states is also low.\footnote{Take a source
that generates $10^5$ pairs per second in average and
consider a 1\,ns detection window, then this probability is
about $10^{-4}$.} Besides, there are techniques that allow
us to eliminate partly multi-photon states. They are based
on the division of the idler beam, used for triggering,
into several detectors. Events with more than one detector
clicks are discarded. This spatial division can be
substituted by time division using one detector behind a
delay loop (\cite{loop}).

The important advantage of a SPDC quasi-single-photon source in
comparison with an attenuated laser is a substantial reduction of
the portion of vacuum contributions, i.e., empty signals.

From the technological point of view these sources seem feasible.
Diode-laser pumped SPDC sources emitting in near-infrared region
can be made compact and robust (\cite{compactDC}).

\subsubsection{Single-photon sources: Color centers}\label{sec_color_centers}

A progressive direction in the research of single-photon sources
is represented by color centers in diamond. Color centers are
defects in a crystal lattice due to impurities and vacancies.
Crystals with such defects can be relatively easily prepared and
are stable. The key advantage of the sources based on color
centers is that they work at room temperatures.

Particularly, nitrogen-vacancy centers in synthetic diamond were
intensively studied (\cite{centers1,centers2,centers3}). These
centers consist of a substitutional nitrogen atom and a vacancy at
an adjacent lattice position. The individual nitrogen atom is
excited by a focused laser beam at 532\,nm. Due to the
fluorescence the atom consequently emits a photon with the
spectrum centered around 690\,nm. The strong anti-bunching is
observed. The weaker point is a broad spectrum of the generated
pulses (nearly 100\,nm). Optical properties of the transmission
medium (absorption, refractive index, etc.) change over such a
large interval of wavelengths. However, recently a new kind of
crystal defect was found that can emit photons at 802\,nm with the
spectral width only about 1\,nm (at room temperature). This color
center consists of a nickel ion surrounded by four nitrogen atoms
in a genuine diamond (\cite{Ni_centers}).

The main problem of single-photon sources based on color
centers is a rather low collection efficiency -- currently
just about 0.1\,\% for bulk crystals. The situation is
slightly better for diamond nanocrystals\footnote{The
subwavelength size of nanocrystals suppresses problems with
the high refraction at the sample interface.} -- currently
over 2\,\% (\cite{centers_QKD1}). The way how to increase
the collection efficiency is to put the crystal into an
optical cavity that suppresses the emission to all other
spatial modes except the preferred one.

There are already first experiments with quantum cryptography
using single photon sources based on nitrogen-vacancy centers
(\cite{centers_QKD1,centers_QKD2}). The QKD was demonstrated in
free space at a distance of 50\,m.

\subsubsection{Single-photon sources: Quantum dots}

Quantum dots are semiconductor nanostructures (``artificial
atoms'') (\cite{dots1,dots2,dots3,dots3a,dots4}). By a suitable
preparation a two- or more-level electronic system can be
obtained. Photon emission comes from recombination of an
electron-hole pair. Electron-hole pairs can be created either by
optical pumping by a pulsed or continuous-wave laser or by an
electric current (\cite{dots_elpump}). Various techniques of
preparation of quantum-dots exist. The usual materials are, e.g.,
GaAs, GaAlAs, or InP.

The wavelength of emitted light is determined mainly by the
material used. Sources operating at telecom wavelengths are
possible (\cite{dots_telecom}). The spectral width of a
generated pulse depends on the number of excited energy
levels and the average number of created electron-hole
pairs.

The main practical drawback of quantum-dot photon sources
is the need of cooling to the liquid-helium temperature.
The latest research promises shift to temperatures about
100\,K (\cite{dots_hightemp}). But the photon-number
distribution of such ``high-temperature'' sources is worse.
The other problem is very low collection efficiency
(usually from $10^{-4}$ to $10^{-3}$). This means that the
probability of obtaining an empty pulse is rather high. The
efficiency can be increased (up to about $10^{-1}$) by
placing the quantum dot into an integrated solid-state
microcavity (\cite{dots_cavity}).

The first demonstration of QKD using a quantum-dot single-photon gun
was done by \cite{dots_QKD}. It operated in free space to a
symbolic distance of one meter.

\subsubsection[Single-photon sources: Atoms and molecules]%
{Single-photon sources: Single atoms and molecules}

Another alternative how to generate single-photon-like
states is to make use of radiative transitions between
electronic levels of a single atom (ion) or molecule.

Single ions caught in a trap and placed inside (or sent into) an
optical cavity where they interact both with the excitation laser
beam and the vacuum field of the cavity (\cite{atoms1, atoms2})
could represent single-photon sources with good properties (with,
e.g., a narrow spectrum and high collection efficiency due to the
presence of the cavity). But practical feasibility of such sources
is still low because of their technological complexity (among
others, high vacuum is needed).

Experiments with single organic-dye molecules are simpler
because the molecules are usually caught in a polymer
matrix (\cite{mol1, mol2,mol3}) or put in a solvent
(\cite{mol4}) and the source is operated in usual
environmental conditions and room temperatures. The photon
statistics of generated states is reported to be good. The
advantage is also a large scope of wavelengths that can be
generated. But the critical problem is a limited stability
of the molecules. Due to the photobleaching even the most
stable dyes survive just a few hours of continuous
excitation.

\subsubsection[Entanglement sources: PDC]{Entanglement source: Spontaneous parametric down conversion}

By spontaneous parametric down conversion (SPDC) one can prepare
photons entangled in energies (wavelengths), momenta (directions),
and/or polarizations. Any of these features can be used for the
purposes of QKD based on Ekert-type protocols (see
Section~\ref{EntPairs}).

In SPDC process one photon from a pump laser is converted,
with a certain (small) probability, into two sub-frequency
photons. The total energy and momentum are conserved
thereat. Since no couple of possible frequencies and wave
vectors of two generated photons is preferred the resulting
quantum state is given as a superposition of all allowed
cases -- it is an entangled state.

SPDC occurs in non-linear optical media. E.g., in crystals
KNbO${}_{3}$, LiIO${}_{3}$, LiNbO${}_{3}$,
$\beta$-BaB${}_{2}$O${}_{4}$, etc. Very perspective SPDC
sources are periodically poled non-linear materials, namely
waveguides in periodically poled lithium niobate
(\cite{PPLN}).

\subsection{Detectors}\label{sec_dets}

\subsubsection{Avalanche photodiodes}\label{sec_APD}

The most widely used detectors in QKD systems with discrete
variables are undoubtedly avalanche photodiodes (APD). In APD a
single photoelectron generated by an impinging photon is
multiplied by a collision ionization. This is because APD
single-photon detectors are operated in a so-called Geiger mode:
On the junction a reverse voltage is applied that exceeds the
breakdown voltage. Thus the impinging photon triggers an avalanche
of thousands of carriers. To reset the detector the avalanche must
be quenched. It could be done by a passive or active way. In the
passive quenching a large resistor is placed in the detector
circuit. It causes the decrease of voltage on APD after the
avalanche starts. in the case of the active quenching the bias
voltage is lowered by an active control circuit. This solution is
faster so that higher repetition rates can be reached (up to
10\,MHz). Another possibility is to work in a so-called gated mode
when the bias voltage is increased above the breakdown voltage
only for a short, well defined period of time.

To detect photons at specific wavelengths different materials of
detector chips are needed. For the visible and near infrared
region (up to 1.1\,$\mu$m) the silicon APD can be used. Nowadays
they are well elaborated. Compact counting modules with integrated
Peltier cooling and active quenching are commercially available
that offer low dark-count rates (below 50 per second) high quantum
efficiencies (up to about 70\,\%) and maximum count rates reaching
10\,MHz. Cooling to temperatures of about $-20^\circ$C is
necessary to keep the numbers of dark counts induced by thermal
noise in a reasonable range. Note that the dark counts, i.e.,
events when the detector sends an impulse even if no photon has
entered it, represent an important factor limiting the operation
range of QKD (see Section~\ref{limitations}).

For telecom wavelengths, 1300\,nm and 1550\,nm, used in fiber
communications, the silicon detectors cannot be applied. For
1300\,nm germanium and InGaAs/InP detectors can be used. Germanium
detectors require cooling to liquid nitrogen temperatures (77\,K).
Typical quantum efficiencies are about 15\,\%, dark-count rates
about $25 \cdot 10^3$ pulses per second (at 77\,K). For 1550\,nm
even germanium detectors cannot be used any more and currently the
only generally available detectors for this wavelength window are
based on InGaAs (on InP substrate). These detectors are now in
common use for both telecom wavelengths. InGaAs detectors must
also be cooled to low temperatures. In practice it can be done
either by three-stage Peltier thermoelectric coolers (down to
about $-60\,{}^\circ$C, i.e., 213\,K) or by compact Stirling
engines (down to about $-100\,{}^\circ$C, i.e., 173\,K). Today's
typical performance of InGaAs APD at 1550\,nm with a Peltier
cooler is as follows: Quantum efficiency about 5--10\,\%,
dark-count rate (in gated mode) about $10^4\,{\mathrm{s}}^{-1}$,
maximal repetition frequency about 100\,kHz--1\,MHz (i.e., dead
time about 1--10\,$\mu$s). And with a Stirling cooler
($-100^\circ$C): Quantum efficiency above 10\,\%, hundreds dark
counts per second (in gated mode), and maximal repetition
frequency about 100\,kHz--1\,MHz. It turns out that the dark-count
rate increases with increasing detection efficiency. It is always
necessary to find a tradeoff between these quantities. As the
number of dark counts increases with temperature, better overall
performance can be achieved at lower temperatures. Also increasing
signal repetition frequency leads to the growth of the number of
dark counts because of the increasing probability of
afterpulses.\footnote{After the avalanche is quenched some charge
carriers may stay trapped on impurities. Their delayed
recombination can lead to so called afterpulses -- unwanted output
impulses of the detector.}

Let us also mention another effect that can play a negative
role in quantum cryptography. When the avalanche is
quenched all charge carriers recombine. It brings the diode
into an insulating state again, a full photodetection cycle
is finished and the diode is ready for the next event.
However, some recombinations are radiative -- this results
in so called backflashes. These dim light pulses propagate
back to the communication channel and they could reveal the
information on Bob's basis setting to an adversary. That is,
they represent a serious side channel and must be carefully
eliminated (blocked) by proper filters (\cite{backflash}).

An interesting possibility to improve the performance of
QKD with APD detectors at telecom wavelengths could be the
combination of parametric frequency up-conversion with
efficient silicon APDs, instead of direct use of InGaAs
APDs. The up-conversion in periodically-poled lithium
niobate can be rather efficient whereas it introduces only
relatively low noise. The overall quantum efficiency in
combination with a silicon APD detector could then be
comparable with the detection efficiency of an InGaAs APD
while the dark-cont rate would be lower (\cite{up-conv}).
This fact could enlarge the operation distance of QKD.

\subsubsection{Quantum dot detectors}

A \emph{Quantum Dot Resonant Tunnelling Diode} is a
semiconductor device with a quantum dot layer encased inside
a resonant tunnelling diode structure (\cite{DotDet}). In
the diode two n-doped GaAs layers are separated by a
double-barrier insulating AlGaAs layer and followed by a
InAs self-assembled quantum dot layer. The resonant tunnel
current through this double-barrier structure is sensitive
to the capture of a hole excited by the photon by one of
the quantum dots in the adjacent dot layer. The capture of
a hole by the dot can switch the magnitude of the current
flowing through the device.

The maximum detection efficiency measured with the device
at 550\,nm was 12\,\%. However, the reasonable dark-count
rate of 4000\,s${}^{-1}$ was achieved with a detection
efficiency of only 5\,\%. The device was cooled to 77\,K.
Measured sample could detect a new photon every 150\,ns. It
corresponds to about 6\,MHz repetition rate
(\cite{DotDet}). But it is mainly limited by external
electronics and the improvement to about 100\,MHz is
expected in a near future.

Note that the detector manufactured from GaAs cannot be
used in the region of telecom wavelengths. Detectors for
these wavelengths have to be built from other materials
like InP.

\subsubsection{Visible Light Photon Counters}

\emph{Visible Light Photon Counters} (VLPC) are
semiconductor detectors consisting of two main layers, an
intrinsic silicon layer and a lightly doped arsenic gain
layer (\cite{VLPC,VLPC99}). When a single photon is
absorbed a single electron-hole pair is created. Due to a
small bias voltage applied across the device, the electron
is accelerated towards the transparent contact on one side
while the hole is accelerated towards the gain region at
the opposite side. Donor electrons in this region are
effectively frozen out in impurity states because the
device is cooled to an operation temperature of about 6\,K.
However, when a hole is accelerated into the gain region it
easily kicks the donor electrons into the conduction band
by impact ionization. Scattered electrons can create
subsequent impact ionization events resulting in avalanche
multiplication.

When a photon is detected, a dead spot of several microns
in diameter is formed on the detector surface, leaving the
rest of the detector available for subsequent detection
events. If more than one photon is incident on the
detector, it will be able to detect all the photons as long
as the probability that multiple photons land on the same
location is small. Therefore these detectors could perform
efficient photon number state detection (photon number
count). However, in practice they can well discriminate
only between zero, one and more photons because of
multiplication noise.

Quantum efficiency of VLPC is about 90\,\% and dark-count
rate about $2 \cdot 10^4$\,s${}^{-1}$ at 543\,nm (at 6\,K).

\subsubsection{Superconducting detectors}

To detect single photons physical processes in
superconductors can also be employed. A few different
principles have been proposed that are now experimentally
tested. All these detectors require cryogenic environment.
The first kind of detector, usually called
\emph{Superconducting Single Photon Detector}, consists of
thin strips of superconducting material, as niobium
nitrate, interconnected to form a meander shaped ``wire''
(\cite{SSPD}). In this ``wire'' the current bias below the
critical current of the material is maintained. An
impinging photon breaks a Cooper pair and generates a
hotspot that forms a resistive potential. The width of the
strips is designed in such a way that the current forced
around the hotspot exceeds the critical current. This
results in the increase of resistance and a voltage signal
indicating the detection of photon. The recent measurements
show that at 1300--1550\,nm the samples have quantum
efficiency up to 10\,\%, dark-count rate about
0.01\,s${}^{-1}$ and counting rate over 2\,GHz
(\cite{SSPD2}). The measurements were done at temperature
2.5\,K (liquid helium).

Another type of superconducting detector is a \emph{Transition
Edge Sensor} (\cite{TES}). These sensors consist of
superconducting thin films electrically biased in the resistive
transition. Their sensitivity is a result of the strong dependence
of resistance on temperature in the transition and the low
specific heat and thermal conductivity of materials at typical
operating temperatures near 100\,mK. The device produces an
electrical signal proportional to the heat produced by the
absorption of a photon. These detectors can even determine the
number of impinging photons, i.e., they can perform a photon
count. Observed efficiency at temperature 125\,mK is about 20\,\%,
dark-count rate about 0.001\,s${}^{-1}$ (\cite{TES}). The newest
results show even a better performance with a quantum efficiency
over 80\,\% at 1550\,nm (\cite{TES2}). Unfortunately, these
detectors are very slow (dead time is about 15\,$\mu$s) because it
is necessary to remove the heat deposited by each photon
(\cite{TES}).

Next possibility is a \emph{Superconducting Tunnel Junction
Detector} (\cite{STJ}). It consists of two superconducting
electrodes separated by an insulating layer forming together a
Josephson junction. To suppress the tunnelling current through the
junction, a magnetic field parallel to the electrodes (parallel to the
tunnel barrier) is applied. Incident photons break Cooper pairs.
It changes the tunnelling rate according to the absorbed energy.
The operating temperature is of the order of hundreds of milikelvins.
These detectors are able to register photons from infrared to
ultraviolet region.

\subsection{Quantum channels}

\subsubsection{Fibers}

The most promising channels for terrestrial QKD are
undoubtedly single-mode optical fibers. The lowest
attenuations of standard telecom fibers are at 1300\,nm
(about 0.35\,dB/km) and at 1550\,nm (about 0.2\,dB/km).
Unfortunately, for these wavelengths standard silicon-based
semiconductor photodetectors cannot be used. In principle,
it is possible to use special fibers and work around
800\,nm, where the efficient detectors are available. But
the attenuation of fibers at these wavelengths is
relatively high, about 2\,dB/km, and such fibers are not
used in an existing infrastructure. Therefore, the
attention is paid to standard telecom fibers and there is
an effort to develop low-noise and efficient detectors for
wavelengths 1300\,nm and 1550\,nm.

The losses in fibers represent one of the two main factors (see
Section~\ref{limitations}) limiting the operation range of QKD
systems (notice that attenuation 0.20\,dB/km means 99\,\% loss
after 100\,km). Other problems are the strong temperature
dependence of some optical properties of fibers, the disturbance
of polarization states of light in fibers due to the geometrical
phase and the birefringence, and the dispersion.

The distortion of polarization is a crucial obstacle for the use
of any kind of polarization encoding of information. Therefore in
fiber-based QKD systems phase-encoding schemes are usually
employed. However, even in such a case the output polarization
state must be under control. Fortunately, if the fiber is fixed
the polarization properties are relatively stable.

Dispersion affects the temporal width of the broad-spectrum
light pulses. Therefore the sources generating broad-band
signals are not well suitable for fiber QKD. Nevertheless,
there is still a possibility to work near the wavelength of
1310\,nm where the silica fibers have zero chromatic
dispersion or to use fibers with special refractive-index
profile which have zero dispersion
shifted near 1550\,nm.

\subsubsection{Free space}

Quantum key distribution can also be accomplished through free
space. The advantage of this approach is that the atmosphere has
very low absorption around the wavelengths 770\,nm and 860\,nm
where relatively efficient and low-noise silicon semiconductor
detectors can be used. Besides, no optical cables have to be
installed. Also, the atmosphere is not birefringent at these
wavelengths and is only weakly dispersive. The disadvantage is
that the free-space communication can be used only in the
line-of-sight distances, no obstacle may be between communicating
parties. There are also other drawbacks: The performance is highly
dependent on the weather, pollution and other atmospheric
conditions. There are huge differences in attenuation for
different kinds of weather. For instance, for wavelengths near
860\,nm the attenuation of clear air can be below 0.2\,dB/km, in
the case of moderate rain it is about 2 -- 10\,dB/km, and in heavy
mist it can exceed 20\,dB/km. Further, up to altitudes about
15--20\,km there are considerable atmospheric turbulences. The
problem is also a spurious influence of the background light,
especially the ambient daylight. Another issue is the beam
divergence. Due to diffraction the diameter of the beam can be
considerably enlarged in large distances. This effect can cause
additional loss if only a part of the beam is captured by the
receiver.

\section{Limitations\label{limitations}}

There are two main technological obstacles that inhibit the wide
spread of quantum key distribution yet: limited operational range
and low transmission rates.

\subsection{Transmission rate}

The key factor limiting the raw-key rate is the detector's
deadtime (i.e., recovery time of the detector). in the case of
avalanche photodiodes (APD) immediately after the detection event
the detector is not ready for other detection. First of all, the
avalanche of charge carriers must be quenched. However, there is
also a problem with the so called afterpulses -- clicks of
detector caused by spontaneous transitions from long-living traps
(levels in a forbidden band) populated by the preceding avalanche.
It is necessary to wait until the carriers leave the detection
(depleted) region. Typical APD dead time is from about hundred
nanoseconds to a microsecond.

The next factor decreasing transmission rate appears if an
attenuated laser is used as a source for QKD. Due to
security requirements (suppression of multi-photon pulses)
the mean photon number per pulse must be fairly below one,
although this leads to a  high vacuum fraction of signals.

Of course, the crucial decrease of transmission rate is due
to losses in the channel.

The rate of distilled key is further decreased by error-correction
and privacy-amplification procedures. The higher the error rate, the
shorter is the distilled key that is obtained from the same amount
of raw key.

\subsection{Limit on the distance}

The maximal distance over which  secure QKD can be established
decreases with increasing losses and increasing detector noise.
The detector dark-count rate is constant (for a given detector and
settings). But the key-rate decreases with increasing distance due
to cumulative losses. So the relative number of erroneous bits
caused by dark counts grows as long as it is so high that
secure QKD is impossible. Standard amplifiers cannot be used as
they would affect the states of photons in a similar manner as
eavesdropping. Present-day technology allows secure operation up
to about 100\,km.

\subsection{Quantum repeaters}

The use of entangled pairs for QKD (see Section \ref{EntPairs})
offers an important advantage. It enables to extend the radius of
secure communication to practically arbitrary distance (at least
in theory). This can be reached by \emph{quantum repeaters}
(\cite{repeaters}). They can do ``distributed error correction''
without revealing any information on the key. The communication
channel is divided into shorter segments each containing a source of
entangled pairs. At the ends of each segment a
\emph{distillation of entanglement} (\cite{distill}) is performed.
It produces a smaller number of ``repaired" highly entangled pairs
from an originally higher number of pairs damaged during
transmission. Individual segments are ``connected'' by means of
an \emph{entanglement swapping} method (\cite{teleport,swapping}). So
finally Alice and Bob possess highly entangled pairs.

\section{Supporting procedures} \label{sec_procedures}

\subsection{Estimation of leaked information}

Real devices like polarizers, fibers, detectors, etc. are
never perfect and noiseless. Therefore we always have to
tolerate a certain amount of errors. However, we cannot be
sure that these errors do not stem from Eve's activity (Eve
could, e.g., replace some noisy part of the system by
better -- less noisy -- one) so we have to attribute all
errors to Eve. Fortunately, from the observed error rate it
is possible to estimate the information leaked to Eve and
then ``shorten" the established key in such a way that
Eve's information on the new, shorter key is arbitrarily
small.

First, Alice and Bob chose randomly a certain number of
transmitted bits and compare them publicly to estimate the
error rate. The higher the number of compared bits is, the
higher is the probability that the actual error probability
does not exceed the estimated value. Assuming the most
general attack allowed by the laws of quantum physics one
can find the boundary of the amount of information, Eve
could get on the key, in dependence on the error rate
caused by the attack. For the simplest intercept-resend
attack described before (assuming non-continuous
eavesdropping) Eve gets an average information per bit
$I=2\epsilon$, where $\epsilon$ is the bit-error rate. Of
course, this attack is not optimal. The limiting (``worst")
values of $I(\epsilon)$ depend both on the protocol and
implementation. These problems will be discussed in more
detail in Section~\ref{Security}.

\subsection[Error correction]{Error correction for classical bit strings}\label{sec_err_cor}

When Alice and Bob create a \emph{sifted key} by sorting
out signals for which Bob has used the ``wrong'' bases,
their key sequences need not be exactly the same. This may
be caused either by an eavesdropping or by
``technological'' noise. Therefore, Alice and Bob must
correct or eliminate the erroneous bits. Here we describe a
simple error-correction procedure proposed by
\cite{ErrCorr1}.

Alice and Bob first agree on a random permutation of the
bit positions in their strings to randomize the location of
errors. Then they partition the permuted strings into
blocks of size $k$ such that single blocks are believed to
be unlikely to contain more than one error (block size is a
function of the expected bit-error rate). For each block,
Alice and Bob compare the block's parity. Blocks with
matching parities are tentatively accepted as correct. If
parities do not agree, the block is subjected to a
bisective search, disclosing further parities of sub-blocks,
until the error is found and corrected.

To remove errors that remained undetected (e.g., because
they occurred in blocks or sub-blocks with an even number
of errors), the random permutation and block parity
disclosure is repeated several more times, with increasing
block sizes. Once Alice and Bob estimate that at most a few
errors remain in the data as a whole, they change the
strategy (at this point, the block parity disclosure
approach becomes much less efficient because it forces
Alice and Bob to reveal at least one parity bit in each block).
Now they publicly choose random subsets of the bit positions
in their entire respective data strings and compare the
parities. If disagreement is found, the bisective search is
undertaken, similar to that described above. The procedure
is repeated several times, each time with a new independent
random subset of bit positions, until no errors is left.
Alice and Bob are now in possession of a string that is
almost certainly shared but only partly secret.

The revealed parity bits represent an additional information
leaked to Eve that must be taken into account. In order to avoid
this leakage of information during the reconciliation process
either the exchanged parity bits must be one-time-pad encrypted or
the information that is additionally made available to the
eavesdropper must be taken into account during the privacy
amplification step.

Other error-correcting (or reconciliation) procedures are
described by \cite{ErrCorr2} (among others the procedure that leak
a minimum amount of information during reconciliation) and by
\cite{cascade}.

Note that the error correction shortens the bit string at least to
a fraction $1 - h(\epsilon)$, where $\epsilon$ is the error rate
and $h(p) = -p \log_2 (p) - (1-p) \log_2 (1-p)$ is the Shannon
entropy. This is the so called Shannon limit. Practical
error-correcting procedures are less efficient and shorten the bit
string even more.

\subsection[Privacy amplification]{Privacy amplification for classical bit strings}\label{sec_priv_amp}

Let us suppose that both Bob and Eve have already made
measurements and they have some classical information on the key
bits sent by Alice.\footnote{If Eve has attacked the transmission
using quantum probes she can wait with measurements on her probes
until Alice and Bob carry out all necessary supporting procedures
and she can then modify her measurements. The procedures described
below are useful even in such a case. More about security issues
can be seen in Section~\ref{Security}.} If Bob has higher
information on the key sent by Alice than Eve [$I(B;A) >
I(E;A)$]\footnote{$I(X;Y)=H(X)+H(Y)-H(X,Y)$ with $H$ being the
Shannon entropy; see Section~\ref{sec_key_rates}.}, then Alice and
Bob can establish a new secret key, such that Eve has
\emph{negligible} information on it, using only \emph{one-way}
communication. First, Alice and Bob have to carry out an
error-correction procedure in order to have the \emph{exactly
same} bit sequences. At that point, Alice and Bob posses identical
strings, but those strings are not completely private. Next, they
proceed with the following algorithm, called privacy amplification
(\cite{Bennett2,ErrCorr1,PrivAmp1}).

Alice, at random, picks $N$ bits, $[X_1, X_2, \dots, X_N]$,
from the sifted key and performs an exclusive OR logic
operation on them (XOR; here we will denote it by
$\oplus$), which finds their sum modulo 2 (in fact she
calculates a parity bit): $[X_1 \oplus X_2 \oplus \dots
\oplus X_N]$. She tells Bob which bits she did the
operation on, but does not share the result. Bob then
carries out the same operation with his bits on the same
positions: $[Y_1 \oplus Y_2 \oplus \dots \oplus Y_N]$ and
keeps the result. As we have supposed that Alice's and
Bob's bit strings are exactly the same ($X_i=Y_i$), Bob's
result must also be the same as Alice's one.

Bob and Alice next replace each $N$-tuple of key bits with
the calculated XOR value (these values represent a new
key). Meanwhile, if Eve, who has many errors in her key,
tries the same operation, it only compounds her mistakes,
thus her information decreases. For example, if Eve knows
the correct value of each bit with a probability
$p=\frac{1}{2} (1+\varepsilon)$ then she will know the
parity bit with the probability $p'=\frac{1}{2}
(1+\varepsilon^N)<p$ when $\varepsilon < 1$.

To put it in a more formal way,  Alice and Bob share an
$n$-bit string $S$, and we suppose that Eve knows at most
$k$ bits of $S$. Alice and Bob wish to compute an $r$-bit
key $K$, where $r < n$, such that Eve's expected
information about $K$ is below some specified bound. To do
so, they must choose a compression function $g: \{0,1\}^n
\to \{0,1\}^r$ and compute $K = g(S)$. The procedure
described above is an example of a good compression
function. It has been shown by \cite{PrivAmp1} that if Eve
knows $k$ deterministic bits of $S$, and Alice and Bob
choose their compression function $g$ at random from the so
called universal class of hash functions, $g: \{0,1\}^n \to
\{0,1\}^r$ where $r = n - k - s$ for some safety parameter
$s \in (0, n-k)$, then Eve's expected information about $K
= g (S)$ is less than or equal to ${2^{-s}}/{\ln 2}$ bits.

It is worth noting that if even a single discrepancy is left
between Alice's and Bob's data after the error correction
procedure, then after privacy amplification their final bit
strings will be nearly completely uncorrelated.

\subsection[Advantage distillation]{Advantage distillation for classical bit strings}\label{sec_adv_dest}

Even if the mutual information on the key of Bob and Alice is
\emph{lower} than the mutual information of Eve and Alice [$I(B;A)
\le I(E;A)$] it may still be possible to establish a secret shared
key by means of a \emph{two-way} classical
communication\footnote{Two-way communication is anyway necessary
for basis announcement in BB84.} (assuming a noiseless and
authenticated classical public channel; \cite{maurer}).

Alice takes an $N$-bit block, $[X_1, X_2, \dots, X_N]$, of the
sifted-key bits, generates a random bit $C$ and makes the
following encoding (here $\oplus$ means XOR again; note that all
bits of the block are XORed with the same bit $C$): $[X_1 \oplus
C, X_2\oplus C, \dots, X_N\oplus C]$. Finally she sends this
encoded block to Bob. Bob then computes $[(X_1 \oplus C) \oplus
Y_1, (X_2\oplus C) \oplus Y_2, \dots, (X_N\oplus C) \oplus Y_N]$,
where $[Y_1, Y_2, \dots, Y_N]$ is his block of the sifted-key bits
corresponding to Alice's block. Bob accepts only if the result
consists of the equal bits, i.e. either $[0,0,\dots,0]$ or
$[1,1,\dots,1]$. In this case he sets either $C'=0$ or $C'=1$,
respectively, as an element of his new key [note that if $X_i=Y_i$
then $(X_i \oplus C) \oplus Y_i=C$]. If Bob's calculation results
in different bits Bob rejects the block.

This procedure is repeated with the other blocks of the sifted key
and other random bits $C$. In other words, Alice and Bob make use
of a repeat code of length $N$ with only two codewords
$[0,0,\dots,0]$ or $[1,1,\dots,1]$. The sequence of random bits
$C$ sent by Alice and accepted by Bob represents a new key
generated by Alice and the sequence of bits $C'$ accepted by Bob
represents a new key received by Bob. In this way, the probability
that Bob accepts erroneously bit $C$ sent by Alice goes down with
increasing $N$ as $\epsilon^N$, where $\epsilon$ is a bit-error
rate in the original sifted key. Eve, on her side, has to use a
majority vote to guess the bit $C$. Hence, Bob's information on
$C$ may be larger than Eve's information even if Bob's information
on Alice's bits $[X_1 , X_2, \dots, X_N]$ is lower than Eve's one.
On the new key the error correction and privacy amplification may
be applied subsequently.

\subsection{Authentication of public discussion}

In practice, the ``auxiliary'' information transmitted
through the open channel during QKD \emph{could} be
modified, as it is difficult to create a physically
unjammable classical channel. For example, Eve can cut both
the quantum and classical channels and pretend to be Bob in
front of Alice. Therefore the authentication of the
messages sent over the open channel is necessary (the
recipient must be able to check that the message has come
from the ``proper" sender and that it has not been
modified). This procedure requires additional ``key''
material to be stored and transmitted. For quantum
cryptography to provide unconditional security, the
procedure used for authentication of public discussion
\emph{must} also be unconditionally secure. Such
authentication algorithms exist (\cite{WC,Stinson}). They
are based, e.g., on the so-called orthogonal arrays. The
length of the authentication password must always be
greater than the length of the authenticated message, but
the authentication tag (the additional information sent
together with the message to verify its origin and
integrity) is relatively short. This authentication tag
itself is one-time pad encrypted to avoid leaking
information on the authentication password to Eve. A small
random sequence of the same length as the authentication
tag, used for its encryption, needs to be renewed after
each QKD transmission (it may be ``refilled'' from the
established key-string). For example, if the cardinality of
the set of authenticated messages is $(p^d-1)/(p-1)$, where
$p$ is a prime and $d \ge 2$ an integer, an authentication
code can be created with $p^d$ keys and $p$ authentication
tags. The deception probability is then $1/p$
(\cite{Stinson,Dusek}).

Clearly, the authentication requires Alice and Bob to meet
each other at the beginning in order to exchange an
authentication password and primary one-time-pad key for
encrypting the authentication tag. After each transmission,
this key is replaced by a new one, obtained from the
transmitted sequence. Therefore, the QKD cryptosystem works
rather as an ``expander" of shared secret information: Some
initial shared secret string is needed but later it can be
arbitrarily expanded.

\section{Security} \label{Security}

It is the goal of QKD to deliver secret keys to the users.
It differs from classical key distribution schemes as in
QKD we can actually prove the security of the final key
under a very limited number of natural assumptions. These
include, for example,  that an eavesdropper cannot have
access to the data inside the devices of Alice and Bob.

In an experimental implementation one cannot demonstrate
directly secure quantum key distribution: security cannot
be measured as such. Security is a theoretical statement
and refers to specific protocols to generate a secret key
from the data we obtain in an experiment. These protocols
depend on observable parameters, such as the error rate,
the mean photon number of the source and the loss rate of
the signals.  So in an experiment, one verifies the model
assumptions of the theoretical security analysis and
demonstrates that one can operate the device such that the
observed parameters allow the generation of a secret key
following the protocol. It is important that the awareness
of this point increases.

Let us have a closer look at the problem of real life
implementations of QKD schemes (see
Section~\ref{sec_experiments}). All devices we are using will be
imperfect to some degree. Moreover, all quantum channels show
imperfections, for example in the form of a polarization mode
dispersion, dephasing in interferometric schemes, and, dominantly,
loss (\cite{gisin02a}). Basic QKD protocols test for the presence
of an eavesdropper by looking for changes in the quantum
mechanical signals. As a result of imperfections we have to face
the situation that Alice and Bob end up with data that deviate
from the ideal ones. Therefore they would have to abort QKD in an
idealized simple protocol that only tests for the presence of an
eavesdropper: we have to assume the worst-case scenario that the
degradation of the data is not due to the channel imperfections,
but might come from an active eavesdropper. The eavesdropper could
be correlated with the data of Alice and Bob, thus having some
information about them. Moreover, in general Alice and Bob do not
even share an error-free bit-string.

It turns out that there are ways to create a secret key despite
these imperfections. For this, Alice and Bob apply some
postprocessing procedures by publicly communicating over a
classical, authenticated channel. Typically, these procedures
include error correction and privacy amplification (see
Section~\ref{sec_procedures}). It is important to know what key
rate can be achieved from the data without compromising security.
The parameters for the public discussion protocols come from the
security proofs. In this section we will give some background to
security proofs and report on the present status for different
protocols.

\subsection{Attacks on ideal protocols}

Before we start to analyze the security of QKD in more
detail, let us have a look at how Eve could actually perform
her eavesdropping activity. From the theory of quantum
mechanical measurements we know that any eavesdropping can
be thought  of as an interaction between a probe and the
signals. Eve can then measure the probe to obtain
information about the signals.

We distinguish three main types of eavesdropping attacks:
\begin{description}
\item[Individual Attack:]
In the individual attack Eve lets each signal interact with
a separate probe. Eve performs then a measurement on each
probe separately after the interaction. This type of attack
is easy to analyze since it does not introduce correlations
between the signals

\item[Collective Attack:]
The collective attack starts as the individual attack, as
each signal interacts with its own independent probe. At
the measurement stage, however, Eve can perform
measurements that act on all probes coherently. We know
from quantum estimation theory that such measurement can in
some cases give more information about the signals than the
individual measurement. For the analysis it is convenient
that also this attack does not introduce correlations
between the signals.

\item[Coherent Attack:]
This is the most general attack which an eavesdropper can
launch on the quantum signals exchanged between Alice and
Bob. Actually, one can assume the worst case scenario that
Eve has access to all signals at the same time. Then the
sequence of signals is described by one high-dimensional
quantum state, on which Eve can perform a measurement via a
single probe. This type of interaction can introduce any
type of correlations, also between subsequent signals, as
seen by Alice and Bob.
\end{description}
\smallskip
Further variations of these attacks can be obtained by
distinguishing whether Eve has to measure her probes before
Alice and Bob continue their protocol, e.g. by exchanging
basis information in the BB84 protocol, or whether she can
delay her measurement until the very end of the protocol
executed by Alice and Bob.

Note that Eve does not necessarily have to measure the
probe to extract information about the key. The secret key
will be used to encrypt a secret, or be used in a different
cryptographic application, which might also use quantum
tools. So Eve might use her probes from the QKD protocol to
attack the subsequent cryptographic application. The
problem whether we can separate the security analysis of
the different steps is known as composability. This has
been addressed recently by \cite{benor04suba} showing that
also in the quantum case the generation of secret key via
QKD can be separated from the use of this key later on.
This is especially important since part of this secret key
will be used to authenticate the public channel of
subsequent QKD exchanges.

Another question is that of the assumptions to which extent an
eavesdropper can exploit imperfections of Alice's and Bob's
devices. As an example, consider single photon detectors: they are
affected by dark counts and have a non-ideal detection efficiency
(see Section~\ref{sec_dets}). In a paranoid picture, we assume
that Eve can exploit even these imperfections. She might reduce or
eliminate dark counts by a suitable pulse sequence inserted into
the optical fiber leading to Bob's detectors. By a change of
wavelength, she might increase the detection efficiency. Clearly,
a precaution against each individual known attempt can be taken,
though it will be hardly possible to list exhaustively all
possible attacks. In a paranoid picture, we are on a safe side
even if Eve could really do all those things. Actually, it turns
out that this paranoid picture is extremely helpful to provide
actual security proofs.

On the other hand, we can hope to protect against
eavesdropping activities that manipulate Bob's detectors.
In that case, the secure key rate will increase clearly.
However, it turns out, that it is technically harder to
provide unconditional security proofs in this scenario.

In the history of QKD, the individual attack played a
crucial role
(\cite{fuchs97a,ekert94,lutkenhaus96,slutsky98a}) since it
has been easy to analyze in conjunction with the
generalized privacy amplification method. However,
presently the individual attack scenario loses its
relevance since methods have been developed to prove
unconditional security, that is, security against coherent
attacks. Actually, it is widely believed that for typical
protocols one needs only to consider collective attacks,
though only recently  steps have been made to prove this
(\cite{renner05a}).

\subsection[Secure-key rates from classical correlations]%
   {Secure key rates from classical three-party
   correlations}\label{sec_key_rates}

A typical, practical QKD protocol consists of two phases:
\begin{description}
\item[Phase I:]
A physical setup generates quantum mechanical signals.
These are distributed and subsequently measured. As a
result, Alice and Bob hold classical data describing their
knowledge about the prepared signals and the obtained
measurement results.

\item[Phase II:]
Alice and Bob use their authenticated classical channel to
talk about their data, for example by sifting their data,
performing error correction and privacy amplification.
\end{description}
\smallskip
The important question is, how exactly to convert the data
obtained in phase I into a secret key in phase II. To
understand this process and its limitation, let us have a
look into the classical world. Also in classical
information theory unconditional security is being
discussed. There the starting point are identically and
independently distributed random variables with a
probability distribution for data of Alice, Bob and Eve,
$P(A,B,E)$. Once one assumes correlations of a given type,
described by $P(A,B,E)$, one can investigate whether public
discussion protocols can turn these data into a secret key.

There are two main results in this context. The first one
is about a lower bound on the achievable  rate. This has
been given by \cite{csiszar78a}. Remember that the Shannon
entropy $H(A)$ of a random variable $A$, which takes values
$a$ with probability $p(a)$, is defined as $H(A)=-\sum_{a
\in A} p(a) \log_2 p(a)$, and the Shannon entropy of a
joint probability distribution is analogously defined as
$H(A,B)=-\sum_{a \in A \choose b \in B} p(a,b) \log_2
p(a,b)$ (\cite{cover91a}). Then the Shannon mutual
information between two parties holding the random
variables $A$ and $B$, respectively, with a joint
probability distribution $p(a,b)$ is then given by
\begin{equation}
I(A;B)=H(A)+H(B)-H(A,B) \; .
\end{equation}
Then the lower bound for the maximal secure-key rate, $R$, is
given (\cite{csiszar78a}) by
\begin{equation}
\label{csiszarkoerner} R \geq \max
\left(I(A;B)-I(A;E),I(A;B)-I(B;E)\right) \; .
\end{equation}
This lower bound can be achieved, if positive,  in the following
way: Alice and Bob perform error correction (see
Section~\ref{sec_err_cor}) via a one-way method, either by Alice
giving error correction information to Bob, or vice versa,
depending on whether the first or second expression in
Eq.~(\ref{csiszarkoerner}) is bigger. If we encode the error
correction information with a one-time pad to avoid leakage of
additional correlations to Eve, then this reduces the effective
key rate by the fraction $1-I(A;B)$ of the original data. In the
second step, Alice and Bob perform privacy amplification,
shortening their key by the fraction $I(A;E)$ or $I(B;E)$,
depending on the chosen communication direction. In total we find
the key rate given on the right hand side of
Eq.~(\ref{csiszarkoerner}).\footnote{Alternatively, one can send
the error correction information unencoded; then the final key is
shortened in privacy amplification giving the same effective
secret key rate (\cite{cachin97a,nl99a}).}

Surprisingly often, we find that this classical lower bound
is also cited and used  in a QKD scenario, where an
optimization over individual attacks is performed to give
bounds on Eve's information about Alice's or Bob's data.
Note, that the use of the Csisz\'ar and K\"orner formula is
restricted to the classical case of independently and
identically distributed random variables. This can only be
justified if we restrict Eve to individual attacks, which
are not necessarily optimal compared to coherent or
collective attack. Additionally, we have to assume that Eve
attacks all signals in precisely the same fashion, and that
she measures the probes of each signal immediately. It is
clear, that the predicted key rates from this procedure can
give a rough feeling of what to expect from a more detailed
security analysis, but it cannot replace it.

The second important result in the classical three-party
situation is due to Maurer (\cite{maurer,maurer-wolf}).
This result gives an upper bound on the extractable secret
key rate for given $P(A,B,E)$. It can be expressed in terms
of the conditional mutual information $I(A;B|E)$, which is
defined as\footnote{The conditional Shannon entropy is
defined as  $H(X|Y) = - \sum_{x \in X, y \in Y} p(y)\,
p(x|y) \log_2 p(x|y)$ with $p(x|y)$ being a conditional
probability.}
\begin{equation}
I(A;B|E)=H(A|E)+H(B|E)-H(A,B|E) \; .
\end{equation}
The formal definition of the upper bound, the
\emph{intrinsic information} is
\begin{equation}
I(A;B\downarrow E) = \min_{E \to \bar{E}}\left[H(A|\bar{E})
 + H(B|\bar{E})-H(A,B|\bar{E})\right]
\end{equation}
where we minimize over all possible mappings from the
random variable $E$ to the random variable $\bar{E}$ [i.e.,
over all possible random distributions $P(A,B,\bar{E})$
consistent with $P(A,B)$]. The intrinsic information
measures how much Bob learns about Alice's data by looking
at his own data after Eve announced her data (or a function
of her data). The bound is then given by
\begin{equation}
\label{Maurerbound} R \leq I(A;B\downarrow E) \; .
\end{equation}
If Bob's data depend only on Eve's announcement, but no
longer on Alice's data, then the intrinsic information
vanishes and we find that no secret key can be generated.
Note that this statement is true for all possible public
discussion protocols Alice and Bob might come up with
(\cite{maurer-wolf}).

By evaluating the lower and upper bounds one finds a wide gap
between them. Actually, there are no protocols known to achieve
the rate of the upper bound. The method of advantage distillation
(see Section~\ref{sec_adv_dest}) taps into the gap
(\cite{maurer}). There are cases where the lower bound is
initially zero, but after the application of an advantage
distillation step the lower bound for the new, conditional,
correlations is positive.

\subsection{Bounds on quantum key distribution}

So far we have been talking about the classical scenario.
There we had to {\em assume} a specific form of the joint
probability distribution $P(A,B,E)$. In quantum mechanics
we can infer from the observations on Alice's and Bob's
side something about the ways Eve might be correlated to
their data, so we are in a stronger position. At the same
time, we have some added complications: Eve is free to
maintain her probes in a quantum mechanical state. We
cannot force her to measure her probe, thus reducing her
probe to classical data. So we cannot directly use quantum
mechanics to consider the class of joint probability
distributions $P(A,B,E)$ that are compatible with the
observations to apply the Csisz\'ar-K\"orner result. Here
we have to find new lines of argumentation to provide the
security statements, including new lower bounds. However,
in one point the classical statements can be directly
applied: the result of Maurer on upper bounds on the key
rates is valid for QKD. Any individual attack compatible
with the observations and quantum mechanics allows us to
derive a valid upper bound (\cite{moroder05suba}). We
obtain this upper bound by choosing a measurement on the
individual probes. This results in a classical probability
distribution $P(A,B,E)$ and subsequently we obtain an upper
bound on the key rate in the quantum case according to
inequality (\ref{Maurerbound}). Other bounds are given e.g.
by the regularized relative entropy of entanglement
(\cite{horodecki03suba,christandl04a}).

This idea allows us to address a question that is important
for experimental quantum key distribution: which types of
correlated data generated by a set-up of Phase I can lead
at all to a secret key via a suitable designed protocol in
Phase II? More specifically, given a  set of signals for
Alice and a  choice of measurement devices for Bob, and
given that one finds some joint probability distribution
$P(A,B)$ for the  signals and measurement results using
some quantum channel under Eve's control: can we at all
generate a secret key from these data? What would be an
upper bound for the data rate we can obtain?

As a (partial) answer it turns out that it is a necessary
condition for generating a secret key from these data that they
cannot be explained as coming from an  entanglement breaking
channel (\cite{curty04a}). Such a channel breaks the entanglement
of an entangled input state by acting on that sub-system of a
bi-partite state which passes through it. It has been shown by
\cite{horodecki03a}, that each entanglement breaking channel can
be represented by a so-called {\em intercept/resend attack} (see
Section~\ref{Int_res_signal_basis}). In this attack Eve performs
some measurement on Alice's incoming signal, transmits the
measurement result over a classical channel and then feeds a new
quantum state into Bob's measurement device which depends only on
Eve's measurement result. If the data cannot be explained in this
way, we say that the data contain quantum correlations. In this
situation it has been shown that the intrinsic information does
not vanish (\cite{acin05a}).

It is easy to see that from data that can be explained as
coming from an entanglement breaking channel we cannot
generate a secret key. Just have a look at the joint
probability distribution of Alice, Bob and Eve, regarding
Alice's signals and Bob's and Eve's measurement results.
This class of channels assures that the joint probability
distribution for Alice and Bob conditioned on Eve factors
as $P(A,B|E)=P(A|E) P(B|E)$. One can insert this into the
definition of the intrinsic information (using $\bar{E}=E$)
and finds quickly that the intrinsic information vanishes,
using $H(A,B|E)=H(A|E)+H(B|E)$. This means that the  upper
bound on the key rate vanishes and no secret key can be
generated. This principle allows us to narrow down the
parameter regimes in which QKD can be successfully
performed at all for specific setups. For specific
protocols, e.g choice of signals and measurement devices,
one can convert the question whether a given set of data
can be explained by an entanglement breaking channel into
the problem of proving the existence of entanglement of a
virtual bi-partite quantum state
(\cite{curty04a,curty05a}). This can be done e.g. using the
idea of entanglement witnesses (\cite{horodecki96a}).

Since general security proofs can be quite complicated, it
makes sense for newly proposed QKD protocols to check first
for which parameter regime of the channel  the upper bound
does not vanish. Note that once we verified the presence of
quantum correlations we only satisfied a necessary
condition for secure QKD, but we still need to provide  a
protocol of Phase II together with a security proof to
achieve QKD. It is not clear whether one can always
generate a secret key once we have quantum correlations.

\subsection{Security proofs}

It is time to show ideas of how to construct protocols in
Phase II which turn the observed correlated data into
secret key. The key requirement in quantum key distribution
is that at the end of such a protocol, the quantum system
in Eve's hand should be uncorrelated with the output of the
protocol: the secret key.

There are several ideas how one can achieve this goal.
Consider a quantum channel which transmits faithfully two
non-orthogonal states. One can show that in this case Eve
cannot have interacted with the signals; more precisely,
starting with a general interaction with a probe and adding
the constraint that the interaction leaves two
non-orthogonal signal states invariant, one can show that
the output of this action is a tensor product between the
probe and the signal states. This guarantees that the probe
cannot be correlated with the signals or Bob's measurement
results: the state of the probe is independent of these
classical data of Alice and Bob.

Clearly, in a realistic noisy channel, we cannot expect to be able
to use this principle directly. However, there is an analogy in
classical information transfer. As we learned from Shannon, one
can use noisy classical channels to transmit classical messages
perfectly. The trick is to use classical error correction codes
that  encode the original message as so called codewords. The
encoded message is sent through the noisy channel. The effect of
the noise on the codewords can be detected and the errors can be
corrected. This mechanism works asymptotically perfect.

Something similar can be done by using Quantum Error
Correction Codes (QECC); \cite{calderbank96a,steane96a}.
Again, the basic idea is to take the non-orthogonal signals
states from the source, to encode them into a longer
sequence of signals that  are transmitted through the
channel, and then to decode the original states
asymptotically error-free. This can be done in principle,
though in this form it would require Alice and Bob to
perform encoding and decoding operations on several
signals, which is beyond our present experimental
capability. Based on this idea, and using earlier results
by Mayers (\cite{mayers96a,mayers01a}), \cite{shor00a}
showed that one can adapt the basic idea of quantum error
correction codes so that the quantum protocol becomes
equivalent to the standard BB84 protocol in which Alice
sends a random sequence of signals and Bob measures them in
a randomly selected basis. In that case, the decoding
operation of the QECC turns into classical error correction
and privacy amplification and no quantum manipulation
capabilities are required.

Let us have a look at this method in more detail. A QECC can
correct errors which are introduced by the channel. The Shor and
Preskill security proof is based on the Calderbank-Shor-Steane
QECC (\cite{calderbank96a,steane96a}) which divides the errors
into bit and  phase errors. That is, without loss of generality,
the channel applies to each signal qubit either an error operator,
the $\sigma_x$ or the $\sigma_z$, or it applies the identity
operator. One encodes the signal qubits into quantum codewords,
e.g. into a larger number of qubits, which are then sent over the
channel. As long as the number of qubits affected by error
operators is sufficiently low, the action of the channel can be
reverted, thanks to the additional structure that is provided by
the codewords. The reversion of the $\sigma_x$ corresponds to the
classical bit error correction. The errors coming from $\sigma_z$
will not be corrected, as we are interested only in the bit values
of the original quantum signals. Instead, one chooses the QECC
structure such that, in principle, one could have corrected the
errors in the quantum domain. This happens by including redundancy
in the signals. Taking out this redundancy is exactly what happens
in the privacy amplification procedure.

We note that one essential step is to estimate the number
of phase and bit errors, since the security hinges on the
fact that one could in principle correct these errors.
Therefore, in fact, it is an essential task to estimate the
number of errors from the observable data. From this
estimation, we can then determine the parameters
characterizing the classical bit error correction and
privacy amplification. It is important to reduce this
estimation problem from the quantum level to the level of
classical estimation theory. In the case of the BB84
protocol and the Shor-Preskill proof this is
straightforward, due to the symmetry. For other protocols
more advanced methods have been developed
(\cite{tamaki03a,koashi04a}).

Let us come to the next principle for security proofs. The
principle exploiting the QECC method uses effectively only
one-way communication. This idea can be extended to two-way
communication, which turns out to tolerate higher noise
levels in the channel. So far, we have been using the idea
that it is sufficient to create an effective perfect
channel between Alice and Bob to guarantee that Eve
decouples from Alice and Bob. Another way to achieve this
goal is to establish maximally entangled states between
Alice and Bob. Once Alice and Bob verify this property,
they can be assured that Eve is decoupled from their
bi-partite states. This is what is commonly referred to as
monogamy of entanglement. Clearly, once we have effective
perfect channels via QECC, we can achieve the distribution
of maximally entangled states. For this, Alice prepares
these states locally and sends one subsystem of each state
to Bob via the effective, perfect, channel. This method can
be generalized in the way that Alice sends the subsystems
via the noisy channel to Bob. The important idea is that
Alice and Bob then perform entanglement distillation to
regain a reduced number of maximally entangled states
(\cite{bennett96b}). This assures that Eve is decoupled
from their states. Actually, the use of one-way QECC is one
method for this, though there are two-way protocols that
can tolerate a higher error threshold. In practical QKD it
is important to find those entanglement distillation
protocols that can be translated again in classical
post-processing of data. An example of this is the protocol
and security proof based on the BB84 protocol by
\cite{gottesman03a} and \cite{chau02a}.

For quite a while it seemed that the security of QKD can be
expressed always as an underlying entanglement purification
protocol. However, recently it has been shown by the
Horodecki family and Oppenheim (\cite{horodecki03suba})
that one can go even further. They showed that one can
create secret keys also from states that are bound
entangled, that is from states that cannot be distilled to
maximally entangled states. The important idea behind their
protocols is that there are certain global unitary
operations acting on their systems only, which cannot
actually be performed by Alice and Bob due to their spacial
separation, but which would turn the bound entangled states
into products of maximally entangled states and some
remaining systems. Again, Eve is then decoupled from the
maximally entangled system. Alice and Bob obtain their
secret key by measuring the maximally entangled state in a
predefined basis. The discussed global unitary operations
now have the property that they leave these measurement
results invariant. So the key data will be the same with or
without applying the unitary operation. Since the key is
secure after application of the global unitary operation of
Alice and Bob, it is also secure without performing this
operation. The security is therefore not based directly
only on the distillability of maximally entangled states.

\subsection{Specific attacks}

Before we turn to the security results for given protocols,
we list a few specific attacks, especially those that are
applicable to realistic implementations of QKD going beyond
the simple qubit picture.

\subsubsection{Intercept-resend attack}

We understand under the intercept-resend attack any attack where
Eve performs a complete measurement on the signals which Alice
sends out. A special version has been introduced already in
Section \ref{Int_res_signal_basis}. Eve then transmits the
classical measurement result and prepares a new quantum state
close to Bob's detection device. In this way, she cuts out all
channel imperfections. As we have seen before, the resulting
correlations will not allow Alice and Bob to create a secret key.
The simplest example is an intercept-resend attack in the BB84
protocol: Eve performs a measurement of the BB84 signals in one of
the signal bases and prepares a state which corresponds to her
measurement result. For example, if she measures in the
horizontal/vertical polarization basis and obtains a vertically
polarized photon, she prepares such a vertical polarized photon
for Bob. Actually, in the sifted key, that is for those signals
where Alice's and Bob's polarization basis agrees, this leads to
an error rate of $25\%$. This error rate is composed of an error
rate of $0\%$ whenever Eve used the same basis as Alice and Bob,
and $50\%$ whenever her basis differs from theirs. It follows,
that for data with more than $25\%$ average error rate QKD cannot
be successfully completed.

\subsubsection{Unambiguous state discrimination attack}

Let us turn to an attack that is a special case of an
intercept-resend attack. It applies whenever the signal states
sent by Alice are linearly  independent. In this case, Eve can
measure the signals with an unambiguous state discrimination (USD)
measurement so that with some probability she learns, without
error, the exact signal, while in the remaining cases she is left
without any information about the signal states (\cite{dusek00a}).
She can now selectively continue her attack. For example, she
might forward a new signal to Bob only in those cases where she
knows the signal for certain, while she might send no signal at
all (corresponding to sending the vacuum state) in the remaining
cases. With this strategy she is able to mimic a lossy channel. As
a result, the data obtained by Alice and Bob show no obvious trace
of eavesdropping whenever Bob obtains a signal. Despite this
absence of visible disturbance of the signal degree of freedom, no
secure key can be created. A typical protocol for which this
problem arises is the variation of the B92 protocol
(\cite{Bennett92}) which uses single photons in non-orthogonal
polarization states together with single-photon detections (see
Section~\ref{sec_B92}). This protocol becomes insecure once the
transmissivity of the channel sinks below a threshold which
depends on the non-orthogonality of the signal states. The
threshold is defined as the transmissivity where the probability
of success of the USD measurement equals the detection probability
for Bob via the lossy channel. In our example, the success
probability of the USD measurement is given as
$P_{\mathrm{USD}}^{\mathrm{succ}}=1-|\langle
\varphi_0|\varphi_1\rangle|$ and Bob obtains the fraction $\eta$
of signals, where $\eta$ is the transmissivity of the channel.
Then we find for the threshold  of the transmissivity the
expression (\cite{tamaki03b})
\begin{equation}
\eta_{\mathrm{thresh}}=1-|\langle
\varphi_0|\varphi_1\rangle| \; .
\end{equation}

\subsubsection{Beam-splitting attack}\label{sec_BS}

The beam-splitting attack is a very natural attack for any optical
implementation of QKD. The reason is that a lossy optical
transmission line is very well described by a model consisting of
an ideal line in which a beam-splitter is inserted which mimics
the loss of the original line. Now Eve gets hold of the signal
emerging from the second output of the beam-splitter, while Bob
obtains the transmitted part. In some protocols, Eve can in these
cases learn a fraction of the signal deterministically
(\cite{ErrCorr1,dusek00a}). This is the case, for example, in
implementations of the BB84 protocol with weak laser pulses
instead of single photons. Alice prepares here weak laser pulses
in the BB84 polarizations such that the signals contain also
multi-photon pulses. The beam-splitter in Eve's attack gives for
some of the signals some, or even all,  photons of a signal pulse
to Eve. She waits until Alice and Bob publicly communicate the
polarization bases of the signals and measurement results. Then
she measures her photons in the correct basis and obtains
deterministically Alice's signals. If also Bob received at least
one photon, then Eve knows deterministically also a bit of the
sifted key (\cite{inamori01suba}). One can show that the secret
key rate is therefore bounded by
\begin{equation}
R \leq p_{\mathrm{exp}}- p_{\mathrm{split}} \; ,
\end{equation}
where $p_{\mathrm{exp}}$ is the probability that a signal enters
the sifted key, and $p_{\mathrm{split}}$ is the joint probability
that Eve obtains at least one photon of the signal \emph{and} that
this signal enters the sifted key. In the case of weak laser
pulses with mean photon number $\mu$, we find
\begin{equation}
R \leq \left(1-e^{-\mu \eta}\right) \left(1- e^{-\mu
(1-\eta)}\right) \; .
\end{equation}
Actually, this upper bound is positive for all values of
the average photon number $\mu$ and of the total
transmissivity $\eta$. It is clear that this attack cannot
be excluded by Alice and Bob by any additional test of the
channel since it represents the physical model of the
channel.

\subsubsection{Photon-number splitting attack} \label{sec_pns}

In the beam-splitting attack the photons of the incoming
signal states are distributed statistically to Eve and Bob.
In principle, Eve could arrange a more effective method
(\cite{Dusek-BSatt,Lutkenhaus,PractQC}). We have seen that
Eve learns an element of the sifted key whenever she
\emph{and} Bob obtain at least one photon. The
beam-splitter, however, sometimes sends all photons of
multi-photon pulses either to Eve or Bob.

The improved eavesdropping attack, called photon-number splitting
attack, starts with Eve performing a quantum non-demolition
measurement of the total photon number of the signals. Whenever
Eve finds a multi-photon signal, she deterministically splits one
photon off, sending the other photons to Bob. Additionally,
whenever she finds a single photon, she either blocks the signal
or she performs a standard eavesdropping method on it and sends it
on to Bob. As we see, errors in the polarization of the signal
arises only by the eavesdropping on the single-photon signals.
Ignoring this effect for the moment, we find again an upper bound
on the possible secret key rate in analogy to the formula for the
beam-splitting attack as (\cite{PractQC})
\begin{equation}
R \leq p_{\mathrm{exp}}-p_{\mathrm{multi}}
\end{equation}
where now $p_{\mathrm{multi}}$ is the joint probability
that Alice sent a multi-photon signal and the signal enters
the sifted key, while $p_{\mathrm{exp}}$ is the total
probability that a signal enters the sifted key. We can
evaluate this bound for a Poissonian photon number
distribution with average photon number $\mu$ and a
single-photon transmissivity $\eta$ for the channel. In
this case we find
\begin{equation}
R \leq (1+\mu)e^{-\mu}-e^{-\mu \eta}
\end{equation}
which is positive only for certain combinations of $\mu$ and
$\eta$. Generally, for given $\mu$ there is a cut-off
transmissivity below which no secure key rate can be generated.
Note that for a realization of this attack it is important that
Eve can suppress signals at will (here some single-photon signals)
without paying any penalty in form of an error rate (see
Sections~\ref{sec_SARG} and \ref{sec_decoy}).

\subsection{Results}

So far we discussed the principles of security proofs and
specific attacks. Next we will summarize results of
complete security analysis as they are known so far. The
results are typically given only in the limit of a large
number of signals, so that all statistical effects of
finite sequences of signals can be neglected.

\subsubsection{Bennett 92 protocol with single photons}

The Bennett protocol of 1992 (B92 protocol) uses only two
non-orthogonal signal states. As discussed before, this
protocol is prone to the USD attack. Nevertheless, it is
possible to achieve unconditional secure key distribution
over lossy channels by adapting the overlap of the input
signal states. This protocol has been analyzed   for
lossless channels (\cite{tamaki03a}) and for lossy channels
(\cite{tamaki04a}). There is no explicit closed formula for
the key rate, for a detailed discussion see the original
publications.

\subsubsection{BB84 protocol with single photons}

The security of the BB84 protocol is well studied
(\cite{mayers96a,mayers01a,shor00a}. Mayers proof
did not make use of random permutations of the signals and
resulted in a secure key rate given by
\begin{equation}
R= 1- h(\epsilon)-h(2\epsilon),
\end{equation}
where $\epsilon$ is the observed error rate and $h(x)$ is the
binary entropy function given by $h(x)=-x \log_2 x -(1-x) \log_2
(1-x)$. The secure rate given by Shor and Preskill is higher, as
they include a random permutation of the signals, so that they
obtain
\begin{equation}
R= 1- 2\, h(\epsilon).
\end{equation}
The cut-off error rate in this scenario is about $11\%$.
However, we know that one can verify quantum correlations
up to $25\%$. \cite{gottesman03a} proposed a two-way
communication protocol in the public discussion part of the
protocol (Phase II) which can come closer to this upper
bound. It has been improved by \cite{chau02a} to tolerate
$20\%$. This is at present the highest known error rate
threshold for the BB84 protocol.

For this protocol, any loss in the channel reduces the
rates only by a prefactor corresponding to the
single-photon transmissivity.

The key rates are given here without the prefactor $1/2$
which would be expected since only in half of the cases the
signal bases of Alice and Bob match. As \cite{efficient}
pointed out, Alice and Bob can choose the probabilities for
the two signal bases asymmetrically. In the limit, they use
basically only one basis, and test only a small number of
signals in the other basis. Though this requires a larger
sampling size, we can nevertheless get rid of the factor
$1/2$ in the rate formulas.

\subsubsection{The 6-state protocol}

The six state protocol can be analyzed in similar fashion to the
BB84 protocol. This has been done by \cite{lo01a} who found the
key rate
\begin{equation}
R= 1+(1-\frac{3 \epsilon}{2})\, \log_2(1-\frac{3 \epsilon}{2})
+\frac{3 \epsilon}{2}\, \log_2 \frac{\epsilon}{2} \; .
\end{equation}
Again, we made use of the idea that one can use the three
bases of the protocol asymmetrically so that we do not have
a prefactor $1/3$.

Also for this protocol there are improved two-way
protocols. The best error threshold found so far is given
by \cite{chau02a} as $27.6\%$.

\subsubsection{BB84 protocol with weak laser pulses}

For practical realizations the BB84 using weak laser pulses
has special importance. The security of this protocol has
been investigated by \cite{inamori01suba}. For this case we
do not only have the key rate for long sequences, but also
the complete analysis for finite key sizes. It extends the
results by Mayers for the single-photon BB84, and therefore
does not use  the random permutation of signals. This
random permutation has been introduced by
\cite{gottesman04a}, so that the final key rate in the long
key limit is given by
\begin{equation}
R = (1-\Delta) - h(\epsilon) - (1-\Delta)\,
h\left(\frac{\epsilon}{1-\Delta}\right),
\end{equation}
where $\Delta$ is the fraction of signals received by Bob
which might have leaked all its signal information to Eve
via a multi-photon process. This fraction is given via the
multi-photon probability of the source,
$p_{\mathrm{multi}}$, and the total signal detection
probability for Bob, $p_{\mathrm{exp}}$, as
\begin{equation}
\Delta = \frac{p_{\mathrm{multi}}}{p_{\mathrm{exp}}}
\end{equation}
This result holds against the most general attack of Eve,
the coherent attack where Eve may delay her measurements.
Moreover, it allows to give reasonable secret key rates
already in the paranoid picture where all of Bob's
detection imperfections (dark counts, detection efficiency)
are ascribed to Eve.

Clearly one can optimize the parameters of the experimental
set-up. By variation of the mean photon number $\mu$ of the
signals we find that one should choose approximately $\mu
\approx \eta$ so that the key rate scales as $R \sim
\eta^2$; $\eta$ is the total transmissivity.

\subsubsection{BB84 with weak laser pulses and decoy states}

The BB84 protocol with weak laser pulses gives a rate of $R
\sim \eta^2$ which is mainly given by the photon-number
splitting attack. One possibility to avoid this attack is
to use the so-called decoy-states
(\cite{H03,LMC05,wang04suba,W04}). Here Alice tests the
channel not only with signals having one average mean
photon number. Instead, she randomly varies the mean photon
number; this she might do with two, three, or many
intensity settings. The idea is that Eve can now no longer
complete the full PNS attack. Of course, she can still
split one photon from each multi-photon pulse, but she can
no longer block the correct number of single-photon signals
for each subset of signals with the same average photon
number. Effectively, this forces Eve back to use the
beam-splitting attack only.

This basic idea is supported by the full security analysis
(\cite{LMC05}), and one finds that the final key rate
scales as $R\sim \eta$, which is a clear improvement of the
performance of these schemes. Indeed now distances of more
than 100 km are possible without giving up a conservative,
paranoid security notion.

\subsubsection{B92 with a strong phase reference pulses}

Another approach to improve the rate of QKD protocols is
the use of coherent states with phase reference. The idea
here is, again,  to make it impossible for Eve to suppress
signals without paying a penalty. The ability to do just
that is what makes the USD attack and the PNS attack so
powerful. This scheme has been analyzed by
\cite{koashi04a}, who confirmed that in this case the
secure key rate scales again as $R\sim \eta$.

\subsection{Side channels and other imperfections} \label{sec_side_ch}

So far we discussed the security assuming that the signals
are prepared exactly as described in the protocol. However,
in physical realizations there might be many imperfections.
For example, the preparation of different signal
polarizations might also affect other degrees of freedom of
the signals, for example  the timing or the spectrum of the
signals. Therefore, by monitoring other than the intended
degrees of freedom Eve might obtain information about the
signal which is not captured in the typical security
analysis. This situation applies also to classical
cryptography where measurable quantities such as power
consumption might help to break classical ciphers.

Other imperfections come into play. Consider the detection
process: typically, we assume that the choices of signals happen
at random. What if Eve can have some information about the basis
or signal choice beforehand, if the detectors show some dependence
of the chosen signal  basis, or if Eve could manipulate the
detectors to some degree? One example is Eve's strategy to apply a
simple intercept-resend attack mimicking Bob's measurement
strategy. Then Eve forwards not only a single photon, but a strong
light pulse in the polarization that corresponds to the
measurement result. If Bob's and Eve's measurement bases agree,
Bob just recovers the signal without error. When the bases
disagree, with almost certainty Bob will find that both of his
single-photon detectors will fire. If Bob  discards these events,
this would open a loophole for Eve to manipulate Bob. For this
reason, Bob has to keep those events, effectively increasing the
error rate since he has to assign a random outcome.

Further, the setting of Bob's measurement basis could be betrayed
by detector backflashes (see Section~\ref{sec_APD}). Eve could
also try to flash to Alice's device and hope to get her setting by
measuring the reflected light. All similar possibilities must be
carefully considered and eliminated.

These questions are currently under investigation. One
finds often the term `Trojan horse attack', as coined by
\cite{lo01a}, for any attack which exploits the
circumstance that Alice's and Bob's devices do  operate not
only on the degree of freedom as specified in the ideal
protocol. It turns out, that many imperfections, once one
has a quantitative bound on them, can be dealt with
(\cite{gottesman04a}). As long as they are small, the
influence on the resulting key rates are small.

\section{Prospects}

It is apparent that quantum cryptography is now ready to
offer efficient and user-friendly systems providing an
unprecedented level of security. While classical methods
are still safe enough for short-lifetime encryption,
quantum cryptography may prove valuable when thinking with
longer prospects. The progress in the development of
quantum computers can play a significant role in speeding
up the increase of the need for QKD in the IT market.
Quantum key distribution can also be well combined with
existing infrastructure. Even QKD with very low bit rate
(hundreds of bits per second) can significantly improve
security of contemporary cryptosystems. It enables, e.g., to
change the secret key for symmetric ciphers like AES
several times per second.

The widespread use of QKD is now restrained mainly due to
the limited operational range (up to about 100\,km). There
are three main technological challenges that can help to
improve this situation: Substantial reduction of noise of
detectors working at wavelengths suitable for fiber
communications (1550\,nm), the development of
ultra-low-attenuation fibers (based, e.g., on photonic
crystals), or the development of quantum repeaters.

Challenging opportunity for future global secure networks
is a long distance quantum communication between Earth and
satellite or between two satellites or satellite and plane
(\cite{satellite1}). The disturbing influence of atmosphere
constraints terrestrial free-space quantum cryptography to
short-range communications. On the other hand in the outer
space and higher levels of atmosphere (above 10~km) only
losses due to beam geometry are important.

\section*{Acknowledgement}

This work was partially supported by the SECOQC project of the EC
(IST-2002-506813), by the Deutsche Forschungsgemeinschaft via the
Emmy-Noether Programme, by the project MSM6198959213 of the
Ministry of Education of the Czech Republic, and by the project
202/05/0486 of the Czech Science Foundation. The authors would
like to thank Marcos Curty, Hauke H\"aseler, and Miroslav Gavenda
for their feedback on the manuscript.

\end{document}